\documentclass[preprint,aps,prc,amsmath,amssymb,showpacs,showkeys,floatfix]{revtex4}

\usepackage{graphicx}
\usepackage{bm}
\usepackage{epstopdf}
\usepackage{float}
\usepackage{booktabs}
\usepackage{dcolumn}
\usepackage{bm}
\usepackage{mathrsfs}
\usepackage{amsmath}
\usepackage{footnote}
\usepackage{accents}
\usepackage{mathtools}
\usepackage{tensor}
\usepackage[titletoc]{appendix}
\usepackage{color}
\usepackage{accents}
\usepackage{multirow}

\begin{document}



\title{The Universe according to DESI results}
\author{Davide Batic}
\email{davide.batic@ku.ac.ae}
\affiliation{
Department of Mathematics, \\
Khalifa University of Science and Technology,
Sas Al Nakhl Campus,
P.O. Box 2533 Abu Dhabi,
United Arab Emirates
}

\author{Sergio Bravo Medina}
\email{sergiobravom@javeriana.edu.co}
\affiliation{
Departamento de F\'isica,\\ Pontificia Universidad Javeriana, Cra.7
No.40-62, Bogot\'a, Colombia
}
\author{Marek Nowakowski}
\email{marek.nowakowski@ictp-saifr.org}
\affiliation{
 ICTP-South American Institute for Fundamental Research,
 Rua Dr. Bento Teobaldo Ferraz 271, 01140-070 S\~ao Paulo, SP, Brazil\\
 and\\
 Departamento de Fisica,  Universidade Federal de S\~ao Paulo,\\
 Unifesp Campus Diadema, - Diadema -  S\~ao Paulo, Brazil
 Rua São Nicolau, 210 \\
CEP: 09913-030 Centro - Diadema -  S\~ao Paulo, Brazil
}

\date{\today}

\begin{abstract}
The recent fit of cosmological parameters by the Dark Energy Spectroscopic Instrument (DESI) collaboration will have a significant impact on our understanding of the universe. Given its importance, we conduct several consistency checks and draw conclusions from the fit. Specifically, we focus on the following key issues relevant to cosmology: (i) the acceleration of the universe's expansion, which, according to the fit, differs over cosmological time compared to the standard cosmological model; (ii) the age of the universe, which appears slightly shorter than the age of the oldest stars; and (iii) the solution of the scale factor, both numerically and in an approximate analytical form.

\end{abstract}

\pacs{ }
\maketitle
\section{Introduction}
It is probably fair to say that mankind's quest to understand the
Universe has been a very long undertaking, especially if we consider
what we might call non-scientific models throughout history
\cite{history}. With the advent of scientific methods, and more
recently with General Relativity \cite{Einstein} and observational
data, this long-standing effort has led cosmologists to accept a model
within the framework of General Relativity, known as the $\Lambda$CDM
model, which is based on the principles of homogeneity and isotropy,
along with the inclusion of a cosmological constant $\Lambda$
\cite{Lambda} ( or its different variations with a decaying
  $\Lambda$ \cite{Lambda2}) and a
yet-to-be-discovered particle that constitutes Cold Dark Matter
\cite{DM, DM2}. The latter could be a stable particle protected by a
symmetry like the R-parity in Supersymmetry \cite{Susy} or unstable,
but long-lived which might leave a fingerprint in the observational
data \cite{Klypin}. Actually, one encounters similar challenges in the
cosmology with unstable neutrinos \cite{neutrinos}. Finally, we
mention also the problem of inflation in the early
evolution of the Universe \cite{Martin}.

We observe that the newly included cosmological constant, which is necessary
to explain the acceleration of cosmic expansion, is not without its
issues \cite{Lambda}. The small value required to explain the
acceleration theoretically contradicts the large contributions it
could receive from zero-point energy in quantum field theory. As a
result, many alternative models to $\Lambda$CDM have been developed
\cite{Sotiriou,DeFelice,Olmo,Myrzakulov,JWu,Katirci,Roshan,Board,Cai,RYang,Capozziello,Heisenberg,Khyllep,Koussour,Jimenez,Guangjie,Shiravand,XHarko,Bahamonde,Obukhov,HeisenbergKuhn,Odintsov,fR2,fR2cosm,HLNO,FRLm}. It is not the purpose of the present article to  review in depth all problems of 
  cosmology and its different models. Therefore  we just leave it to
  the reader to consult the many references and focus on the topic of
  the paper: to draw consequences from recent cosmological observations.

From the observational side, the so-called Hubble tension \cite{Verde,
  HubbleTension}, i.e., the measurement of differing Hubble constants
that are not compatible with each other casts some doubts on the
concordance cosmological model, namely, the $\Lambda$CDM
model. Another important development that could challenge the validity
of $\Lambda$CDM is the early release of new data by the DESI
collaboration \cite{DESI}, which presented an epoch-dependent fit to
the equation of state. This significant departure from the constant
ratio of Dark Matter pressure to its density, as encountered in
$\Lambda$CDM, could be the harbinger of a new understanding of the
Universe. It is therefore logical to draw some conclusions from the
DESI data. This has been partially addressed in
\cite{QuintessenceDESI}, which interpreted the data within the context
of a Quintessence model. In this article, we focus on topics not
covered in \cite{QuintessenceDESI}. One of the questions we address in
relation to the DESI results is the acceleration of the expansion. The
current acceleration of the Universe has dominated cosmology in recent
decades, and its status has not changed significantly since the Nobel
Prize in 2011 \cite{U1,U2}. Another equally important issue is the lifetime of the
Universe, especially in comparison with the lifetimes of the oldest
objects we observe. Simply put, the ages of old stars and galaxies
cannot exceed the lifetime of the Universe (or comes too close to it), making this a powerful constraint \cite{our}.

The article is structured as follows: In Section II, we revisit the
key features of the $\Lambda$CDM model, providing analytical solutions
that will serve as a benchmark for comparison with the Quintessence
(scalar field) model discussed in Section III. Within this model, we incorporate DESI data to explore its implications on the Universe's acceleration and lifetime. Additionally, we present numerical solutions for the scale factor and the Hubble function, supplemented by analytical approximations to provide a comprehensive understanding.

\section{The Universe with a cosmological constant}\label{Sec2}
It is useful to outline some basic features of the current concordance model of cosmology ($\Lambda$CDM), which is based on Einstein's General Relativity \cite{Einstein} and employs the Robertson-Walker metric \cite{FRW}. This model includes a positive cosmological constant \cite{Lambda} as Dark Energy and incorporates Dark Matter \cite{DM} into the standard matter density. This overview will serve as a foundation for exploring new models and as a basis for comparison between these models.

From the Friedmann equations with a cosmological constant $\Lambda$ in the flat ($k=0$) FLRW (Friedmann-Lemaître-Robertson-Walker) metric, we have
\begin{eqnarray}
H^{2}&=&\left( \frac{\dot{a}}{a}\right)^{2}=\frac{8\pi G}{3}\rho+\frac{\Lambda}{3}=\frac{8\pi G}{3}\left(\rho+\frac{\Lambda}{8\pi G} \right),\\
\frac{\ddot{a}}{a}&=&-\frac{4\pi G}{3}(\rho + 3P)+\frac{\Lambda}{3}=-\frac{4\pi G}{3}\left[ \left( \rho+\frac{\Lambda}{8\pi G}\right) + 3 \left(P-\frac{\Lambda}{8\pi G} \right)\right],
\end{eqnarray}
where the following identification is made
\begin{equation}
\Lambda=8\pi G \rho_{\rm vac}=-8\pi G P_{\rm vac}.
\end{equation}
By defining $\rho_{\rm tot}= \rho + \rho_{\rm vac}$, and $P_{\rm tot}=P+P_{\rm vac}$,  we can rewrite the Friedmann equations as
\begin{eqnarray} 
H^{2}=\left( \frac{\dot{a}}{a}\right)^{2}=\frac{8\pi G}{3}\rho_{\rm
  tot},\label{F1}\\
\frac{\ddot{a}}{a}=-\frac{4\pi G}{3}(\rho_{\rm tot} + 3P_{\rm tot}).\label{F2}
\end{eqnarray}
We remind the reader that defining $\Omega_m(t)=\rho(t)/\rho_{crit}$
with $\rho_{crit}(t)=3H_0^2/8\pi G$  and $\Omega_{\Lambda}(t) =\rho_{vac}/\rho_{crit}(t)$,  equation \eqref{F1} takes the form
\begin{equation} \label{omega}
  \Omega_{\Lambda}(t)+\Omega_m(t)=1
\end{equation}
Whenever we refer to these definitions at the present time $t=t_0$, we will simply write $\rho_{crit}$, $\Omega_{\Lambda}$ and $\Omega_m$.
We note that for each energy density and pressure, we can define an
equation of state \cite{DEAU} as
\begin{equation}\label{DEA}
w_{i}\equiv \frac{P_{i}}{\rho_{i}}.
\end{equation} 
In the cosmological model with $\Lambda$, we have $w_{\Lambda}=P_{\rm vac}/\rho_{\rm vac}=-1$, which is the equation of state for standard Dark Energy. Moreover, we set an equation of state for the matter energy density as follows
\begin{equation}
P=w\rho\equiv (\gamma -1)\rho,
\end{equation}
where we have defined $\gamma=w+1$ for matter to avoid confusion with $w_{\Lambda}$. Next we combine the Friedmann equations above, namely (\ref{F1}) and (\ref{F2}), to arrive at an equation in Riccati form
\begin{equation} \label{Riccati}
\dot{H}=-\frac{3\gamma}{2}H^{2}+\frac{\gamma\Lambda}{2},
\end{equation}
where $\Lambda$ can be positive or negative. If we introduce dimensionless variables $\xi=\sqrt{|\Lambda|}t$ and $\omega(\xi)=H(t(\xi))/\sqrt{|\Lambda|}$, then the above differential equation takes the form
\begin{equation}
\frac{d\omega}{d\xi}=f(\omega),\quad
f(\omega)=\left\{ \begin{array}{ll}
         +\frac{\gamma}{2}\left(1-3\omega^2\right) & \mbox{if $\Lambda > 0$},\\
         -\frac{\gamma}{2}\left(1+3\omega^2\right) & \mbox{if $\Lambda< 0$}.
\end{array} \right..
\end{equation}

This differential equation can be solved. For instance, by disregarding the solution that yields a negative Hubble parameter and considering $\Lambda<0$,  we obtain
\begin{equation}
H(t)= \sqrt{-\frac{\Lambda}{3}}\tan{(\alpha-\beta t)},\quad
\alpha=\beta t_0+\tan^{-1}{\left(\sqrt{-\frac{3}{\Lambda}}H_0\right)},\quad\beta=\frac{\gamma\sqrt{-3\Lambda}}{2},
\end{equation}
where we have used the initial condition $H_0=H(t_0)$. We can integrate this with the initial condition $a(t_0)=1$, resulting in
\begin{equation}
a(t)=a_0\left[\frac{\cos{(\alpha-\beta t)}}{\cos{(\alpha-\beta t_0)}}\right]^\frac{2}{3\gamma},
\end{equation}
which gives the re-collapsing solution for $t \in
[\frac{\alpha -\pi/2}{\beta}, \frac{\alpha +\pi/2}{\beta}]$. When we solve the same equation for $\Lambda>0$, we encounter two distinct cases. For $0<H <\sqrt{\Lambda/3}$ and with the initial condition $H(t_0)=H_0$, the solution is
\begin{equation} \label{solutionH}
H(t)= \sqrt{\frac{\Lambda}{3}}
\tanh{\left(\epsilon t+\delta\right)},\quad
\epsilon=\frac{\gamma}{2}\sqrt{3\Lambda},\quad
\delta=\tanh^{-1}{\left(\sqrt{\frac{3}{\Lambda}} H_0\right)}-\epsilon t_0.
\end{equation}
With the initial condition $a(t_0)=1$, the scale factor becomes
\begin{equation} \label{solutiona}
a(t)=\left[\frac{\cosh{(\epsilon t+\delta)}}{\cosh{(\epsilon t_0+\delta)}}\right]^\frac{2}{3\gamma},
\end{equation}
which is an unphysical solution since the density becomes negative. Specifically, equation \eqref{omega} now requires $\Omega_{\Lambda}<1$, but $H <\sqrt{\Lambda/3}$ leads to $\Omega_{\Lambda}>3$. The physically correct solution, for
\begin{equation} \label{ineq1}
    H > \sqrt{\Lambda/3},
\end{equation}
is obtained by replacing $\tanh$ with $\coth$ in (\ref{solutionH}) and $\cosh$ with $\sinh$ in (\ref{solutiona}). After fixing the integration constant using the initial condition $H_0=H(t_0)$, the solution can be expressed as
\begin{equation} \label{solutionH2}
H(t)= \sqrt{\frac{\Lambda}{3}}
\coth{(\mu t+\nu)},\quad
\mu=\frac{\gamma\sqrt{3\Lambda}}{2},\quad
\nu=\frac{1}{2}\ln\left(\frac{1+\sqrt{\Omega_{\Lambda}}}{1-\sqrt{\Omega_{\Lambda}}}\right)-\mu t_0,
\end{equation}
where $\Omega_{\Lambda}=\rho_{\rm vac}/\rho_{\rm crit}$, with $\rho_{crit}=\frac{3H_0^2}{8\pi G}$.
The scale factor can then be calculated as
\begin{equation} \label{solutiona2}
a(t)=\left(\frac{1-\Omega_{\Lambda}}{\Omega_{\Lambda}}\right)^{\frac{1}{3\gamma}}
\sinh^{\frac{2}{3\gamma}}{(\mu t+\nu)}.
\end{equation}
If we let $T$ be the time at which $a(T)=0$, then the lifetime of the universe is given by
\begin{equation} \label{lifetime2}
T_{\rm Univ}=t_0-T=\frac{1}{3\sqrt{\Omega_{\Lambda}}}H_0^{-1}
\ln\frac{1+\sqrt{\Omega_{\Lambda}}}{1-\sqrt{\Omega_{\Lambda}}}\,\,\, {\rm Gyr}=9.7777 h_0^{-1}
\frac{1}{3\sqrt{\Omega_{\Lambda}}}
\ln\frac{1+\sqrt{\Omega_{\Lambda}}}{1-\sqrt{\Omega_{\Lambda}}}\,\,\, {\rm Gyr},
\end{equation}
where we have assumed that the duration of the radiation period ($\gamma=4/3$)
is negligible compared to the dust epoch ($\gamma=1$). \cite{KT} derived a similar formula without using explicit solutions. Specifically, their lifetime formula is $T_{\rm Univ}=(2/3)H_0^{-1}\Omega_{\Lambda}^{-1/2}\ln[(1+\Omega_{\Lambda}^{1/2})/(1-\Omega_{\Lambda})^{1/2}]$. It is gratifying to see that after some algebraic manipulations, both expressions are
identical. For $h_0=0.7$ and $\Omega_{\Lambda}=0.73$, the lifetime comes out to be 13.866 Gyr. It is also worth mentioning that the solution found in \cite{Aldrovandi} for $\gamma=1$ is a special case of our more general expression \eqref{solutionH2} with $c/L=\sqrt{\Lambda/3}$ and $H(0)=\cosh{\nu}$.\\
The current accelerated stage of the Universe imposes a condition if
we take $\ddot{a}>0$ from equation (\ref{F2}), with $P=0$, $P_{\rm
  vac}=-\rho_{\rm vac}$, and $\rho(t_{0})=\rho_{0}$. This implies, at
the present epoch
\begin{equation} \label{ineq2}
\rho_{0}+\rho_{\rm vac}+3P_{\rm vac}<0, \quad \Rightarrow \quad \rho_{0}<2\rho_{\rm vac}.
\end{equation}

The value of the matter density in \cite{Cosmographic}
is $\rho_{0}=(0.285\pm0.012)\rho_{\rm crit}$, while in \cite{DESI}
it is given as $(0.295\pm0.015)\rho_{\rm crit}$, 
which includes both baryonic matter and Dark Matter. This sets a constraint on $\rho_{\rm vac}$, at least for the current state of accelerated expansion. The generalization to an arbitrary time (represented here by the scale factor) is $\rho_0/a^3 < 2\rho_{vac}$, or, in other words, provided that
\begin{equation} \label{ineq2.1}
  a^3>\frac{\Omega_m}{2(1-\Omega_m)}.
\end{equation}

One can easily recognize, from a phenomenological point of view, how the Friedmann equations can be generalized to 
\begin{eqnarray}
H^{2}&=&\frac{8\pi G}{3}\left( \rho+\rho_{DE} \right),\label{F1DE}\\
\frac{\ddot{a}}{a}&=&-\frac{4\pi G}{3}(\rho + \rho_{DE}+3(P+P_{DE})),\label{F2DE}
\end{eqnarray}
together with \eqref{DEA}.  Assuming that observations 
yield $w_{DE}=\frac{P_{DE}}{\rho_{DE}}=-1$, this would confirm the standard
cosmological model with $\Lambda$.  Following the early release of
data \cite{DESI}, the DESI fit can be summarized as
follows. Here, $w$ refers to $w=P_{DE}/\rho_{DE}$. The flat $w$CDM with a constant state parameter for Dark Energy is given by
\begin{equation} \label{wconst}
w = -0.99 ^{+0.062}_{-0.054},
\end{equation}
or the flat $w_{0}w_{a}$CDM model, where the state parameter depends on the scale factor $a$ \cite{wa1, wa2}, is expressed as
\begin{equation} \label{eosDesi}
w(a)=w_{0}+w_{a}(1-a),
\end{equation}
with the values from DESI BAO set as
\begin{equation} \label{BAO}
w_{0}=-0.55^{+0.39}_{-0.21}, \qquad w_{a}<-1.32,
\end{equation} \label{BAOCMB}
or when combined with CMB results as
\begin{equation}
w_{0}=-0.45^{+0.34}_{-0.21}, \qquad w_{a}=-1.79^{+0.48}_{-1.0}.
\end{equation}

At this point, it is appropriate to compare the lifetimes of the
Universe derived from Planck data with those obtained using the DESI fit \cite{DESI} (as given in equation \eqref{wconst}), along with the measured Hubble constant. To this end, we present the lifetime calculation using Planck 2018 data \cite{Planck} in Table~
\ref{LifetimeTablePlanck}. We do the same calculation in Table~\ref{LifetimeTableDESI1} using $H_{0}$ from DESI BAO+CMB data and $\Omega_{\Lambda}$ from DESI BAO data, while in Table~\ref{LifetimeTableDESI} we use values coming from DESI+CMB data. As we will discuss in Section~\ref{4D}, the lifetime based on the DESI could be approaching the edge of the allowed limit. 

\begin{table}[ht]
\centering
\begin{tabular}{|c|c|}
\hline
\multicolumn{2}{|c|}{\textbf{Planck Data}} \\ \hline
\textbf{Parameter} & \textbf{Value} \\ \hline
$H_{0}$ & $(67.36 \pm 0.54)$ km s$^{-1}$ Mpc$^{-1}$ \\ \hline
$H_{0}^{-1}$ & $(14.516 \pm 0.1)$ Gyr \\ \hline
$\Omega_{\Lambda}$ & $0.6847 \pm 0.0073$ \\ \hline
$T_{\rm Univ}$ & $(13.8 \pm 0.1)$ Gyr \\ \hline
\end{tabular}
\caption{Estimated lifetime of the Universe based on $\Lambda$CDM using Planck data.}
\label{LifetimeTablePlanck}
\end{table}

\begin{table}[ht]
\centering
\begin{tabular}{|c|c|}
\hline
\multicolumn{2}{|c|}{\textbf{DESI Data (DESI+BAO+CMB)}} \\ \hline
\textbf{Parameter} & \textbf{Value} \\ \hline
$H_{0}$ & $(68.3 \pm 1.1)$ km s$^{-1}$ Mpc$^{-1}$ \\ \hline
$H_{0}^{-1}$ & $(14.316 \pm 0.2)$ Gyr \\ \hline
$\Omega_{\Lambda}$ & $0.651^{+0.068}_{-0.057}$ \\ \hline
$T_{\rm Univ}$ & $(13.2^{+0.8}_{-0.6})$ Gyr \\ \hline
\end{tabular}
\caption{Estimated lifetime of the Universe based on $\Lambda$CDM using DESI data.}
\label{LifetimeTableDESI1}
\end{table}

\begin{table}[ht]
\centering
\begin{tabular}{|c|c|}
\hline
\multicolumn{2}{|c|}{\textbf{DESI Data (DESI+CMB)}} \\ \hline
\textbf{Parameter} & \textbf{Value} \\ \hline
$H_{0}$ & $(67.97 \pm 0.38)$ km s$^{-1}$ Mpc$^{-1}$ \\ \hline
$H_{0}^{-1}$ & $(14.386 \pm 0.008)$ Gyr \\ \hline
$\Omega_{\Lambda}$ & $0.6931\pm 0.005$ \\ \hline
$T_{\rm Univ}$ & $(13.78\pm0.06)$ Gyr \\ \hline
\end{tabular}
\caption{Estimated lifetime of the Universe based on $\Lambda$CDM using DESI data.}
\label{LifetimeTableDESI}
\end{table}

\section{The Quintessence model}
It remains to present a concrete realization of the DE (Dark Energy)-model presented in (\ref{F1DE}) and (\ref{F2DE}). This can be achieved within the framework of the so-called Quintessence models \cite{QuintessenceReview}, where a scalar field $\phi$ is incorporated into the Einstein-Hilbert action
\begin{equation}
S=\int d^{4}x \sqrt{-g} \left[ \frac{1}{16\pi G} R -\frac{1}{2}g^{\mu\nu}\partial_{\mu}\phi\partial_{\nu}\phi - V(\phi)\right] + S_{m},
\end{equation}
as has been partly discussed in\cite{QuintessenceDESI}.
The identification of the pressure and energy density of the scalar field, along with its equation of state parameter, is given by
\begin{equation}
P_{\phi}=\frac{\dot{\phi}^{2}}{2}-V(\phi), \qquad \rho_{\phi}=\frac{\dot{\phi}^{2}}{2}+V(\phi), \qquad w=\frac{P_{\phi}}{\rho_{\phi}}=\frac{\frac{\dot{\phi}^{2}}{2}-V(\phi)}{\frac{\dot{\phi}^{2}}{2}+V(\phi)}.
\end{equation}
It is worth noting that the constant equation of state parameter for
quintessence is constrained by the equations of the model
\cite{Constantw}.  Therefore, the choice of the potential $V(\phi)$ should, in principle, be equivalent to specifying an equation of state.

Now, from the variation of the action in the FLRW metric, we obtain the following equation for the field
\begin{equation} \label{eqPhi}
\ddot{\phi}+3H\dot{\phi}+V_{,\phi}=0,
\end{equation}
which is equivalent to a continuity equation of the form
\begin{equation} \label{conPhi}
\dot{\rho}_{\phi}+3H(\rho_{\phi}+P_{\phi})=0.
\end{equation}
The Friedmann equations now become (assuming $k=0$)
\begin{eqnarray}
H^{2}&=&\frac{8\pi G}{3}\left( \rho_{\phi}+\rho \right),\label{F1D}\\
\frac{\ddot{a}}{a}&=&-\frac{4\pi G}{3}(\rho + \rho_{\phi}+3(P+P_{\phi})).\label{F2D}
\end{eqnarray}

In the above equations, we can refer to the sum of the densities as 
$\rho_{tot}$ and the sum of the pressures as $P_{tot}$. Differentiating
the first equation and using both equations, we obtain the total conservation
law
\begin{equation} \label{conTotal}
  \dot{\rho}_{tot}+3H(\rho_{tot}+P_{tot})=0.
\end{equation}
The conservation law and the continuity equation for $\rho_{\phi}$ and $P_{\phi}$ imply
\begin{equation} \label{conMatter}
    \dot{\rho}+3H(\rho+P)=0,
\end{equation}
which, with the equation of state $P=(\gamma-1)\rho$, gives a solution for $\rho(a)$ as
\begin{equation}\label{rho}
  \rho(a)=\rho_{0}a^{-3\gamma}.
\end{equation}
The Quintessence model also encompasses another well-known class of models: modified gravity, which is described by a Lagrangian of the $f(R)$ type. In $f(R)$ theories of gravity, the standard Lagrangian term associated with the Ricci scalar $R$ is replaced by an arbitrary function of $R$, denoted as $f(R)$. This modification leads to an action that can be expressed as \cite{FofR}.
\begin{equation}\label{FofR}
S_{f(R)}=\int d^{4}x \sqrt{-g}\left[\frac{1}{16\pi G}f(R) \right]+S_{m}.
\end{equation}
If a conformal transformation is applied to the metric in the following form
\begin{equation}
g_{\mu\nu}(x)\rightarrow \tilde{g}_{\mu\nu}(x)=\Omega^{2}(x)g_{\mu\nu}(x),
\end{equation}
with the identification
\begin{equation}
\Omega^{2}=\partial_{R}f(R)=F(R),
\end{equation} 
then the geometric part of the action can be rewritten as \cite{Starobinsky,fRscalar,fRscalar2}
\begin{equation}
S=\int d^{4}x \sqrt{-\tilde{g}}\frac{1}{16\pi G}\tilde{R}-\int d^{4}x\sqrt{-\tilde{g}}\left[\frac{1}{2}\tilde{g}^{\mu\nu}\partial_{\mu}\phi\partial_{\nu}\phi+V(\phi) \right],
\end{equation} 
where $\tilde{R}$ is the Ricci scalar computed for the new metric, and the scalar field and potential are identified as
\begin{equation}
\phi=\sqrt{\frac{3}{16\pi G}}\ln F (R),\quad
V(\phi)=\frac{1}{16\pi G}\frac{FR-f}{F^{2}}.
\end{equation}
Thus, in principle, an identification can be made between an $f(R)$ model and the quintessence model.

In passing, we make two observations. First, the equation corresponding
to the Riccati  equation in (\ref{Riccati})  is ($\gamma=1$)
\begin{equation} \label{noRiccati}
\dot{H}=-\frac{3}{2}H^{2}-\frac{4\pi G}{3}\left[
  3(w_{0}+w_{a}(a-1)) \right]\rho_{\phi},
\end{equation}
which is no longer in Riccati form.  Choosing in the above $w_a=0$,
$w_0=-1$ and $\rho_{\phi}=\rho_{vac}$ equation (\ref{noRiccati}) reduces
to (\ref{Riccati}). 
Secondly, we can
interchangeably use the subscripts $DE$ and $\phi$, identifying
$\rho_{DE}=\rho_{\phi}$ and $P_{DE}=P_{\phi}$.

\subsection{DESI $w_{0}w_{a}$CDM in Quintessence form}
We can apply the Quintessence model by setting $w$ in the form used by DESI, specifically
\begin{equation} \label{eosPhi2}
\frac{P_{\phi}}{\rho_{\phi}}=w_{0}+w_{a}(1-a).
\end{equation}
Substituting this into the continuity equation yields
\begin{equation} 
\dot{\rho_{\phi}}=-3H(1+w_{0}+w_{a}(1-a))\rho_{\phi},
\end{equation}
as demonstrated in \cite{QuintessenceDESI}. The density can then be
solved in terms of the scale factor $a$ as
\begin{equation} \label{rhoPhi}
\rho_{\phi} (a)=\rho_{\phi,0} a^{-3(1+w_{o}+w_{a})}e^{3w_{a}(a-1)},
\end{equation}
where $a_{0}=1$ and $\rho_{\phi,0}$ is the current value of the Dark Energy density. This expression describes a decreasing function of the scale factor 
$a$. When combining both solutions for the densities, we obtain
\begin{equation}
\rho_{\rm tot}(a)=\rho(a)+\rho_{\phi}(a)=\rho_{0}a^{-3\gamma}+\rho_{\phi,0} a^{-3(1+w_{o}+w_{a})}e^{3w_{a}(a-1)}.
\end{equation}
We note that $\rho_{\phi,0}$ has yet to be determined, but it can be obtained  from the Friedmann equation (\ref{F1}) at $t=0$
\begin{equation}
H_{0}^{2}=\frac{8\pi G}{3}(\rho_{0}+\rho_{\phi,0}),\qquad \Rightarrow \qquad \rho_{\phi,0}=\rho_{\rm crit}-\rho_{0},
\end{equation}
where $\rho_{\rm crit}$ and $\rho_0$ are known values. The density $\rho(a)$ can be substituted back into equation (\ref{F1}) and used to solve for $a(t)$ as follows
\begin{equation}\label{equa}
\frac{1}{a}\frac{da}{dt}=\pm \sqrt{\frac{8\pi G}{3}\rho_{\rm tot}(a)},
\end{equation}
from which we find that
\begin{equation}
dt=\pm \frac{da}{a\sqrt{\frac{8\pi G}{3}\rho_{\rm tot}(a)}}=\pm \frac{da}{a\sqrt{\frac{8\pi G}{3}\left(\rho_{0}a^{-3\gamma}+\rho_{\phi,0} a^{-3(1+w_{o}+w_{a})}e^{3w_{a}(a-1)} \right)}},
\end{equation}
where the positive sign corresponds to an expanding universe.
If the solution to the integral is invertible, we can obtain a closed form for $a(t)$ from
\begin{equation}
t-t_{0}=\pm\int_{a=a_{0}}^{a} \frac{da'}{a'\sqrt{f(a')}},\quad
f(a)=\frac{8\pi G}{3}\left(\rho_{0}a^{-3\gamma}+\rho_{\phi,0}
  a^{-3(1+w_{o}+w_{a})}e^{3w_{a}(a-1)} \right).
\end{equation}
As a consistency check, we note that the density $\rho_{tot}(a)$ would have
no singularity. By taking $\gamma=1$ (dust) and $1+w_{0}+w_{a}\simeq
-1.24$ , we find that $\rho_{tot}(a)$ would be zero if
\begin{equation}
\rho_{0} + (\rho_{\rm crit}-\rho_{0})a^{3.72}e^{-3w_{a}}e^{3w_{a}a}=0.
\end{equation}
Since $\rho_{0}$ is positive, the only way for this equation to be zero
would be if $\rho_{\rm crit}-\rho_{0}<0$, or equivalently, $\rho_{\rm
  crit}<\rho_{0}$. However, observations indicate that $\rho_0 <
\rho_{crit}$, which implies that the  density $\rho(a)$ does not exhibit any singular behaviour. This result would also hold for $\gamma=4/3$ or for any values of $w_{0}$ or $w_{a}$.

\subsection{The acceleration}
The accelerated expansion of the Universe has been a crucial cornerstone of cosmology over the past few decades. Observations of type II supernovae established this fact, which even led to the awarding of the Nobel Prize in 2011, confirming that indeed $\ddot{a}>0$ \cite{U1,U2}. This discovery spurred many cosmologists to develop new cosmological models. The simplest of these, still within the framework of Einstein's gravity, involves introducing a positive
cosmological constant. Given the recent DESI data, it is of significant interest to further investigate this matter. 

Imposing $\ddot{a} >0$ at the present epoch $a=a_{0}=1$, the relevant Friedmann equation becomes
\begin{equation}
-\frac{4\pi G}{3}\left(\rho_{0}a_{0}^{-3}+\rho_{\phi}(a_{0})+3P_{\phi},(a_{0})\right)>0
\end{equation}
which, using $\rho_{\phi,0}=\rho_{\rm crit}-\rho_{0}$, simplifies to
\begin{equation}
(3w_{0}+1)\rho_{\rm crit}<3w_{0}\rho_{0}.
\end{equation}
Given that $\rho_{0}=(0.285\pm0.012)\rho_{\rm crit}$ and $w_{0}<0$, this inequality leads to
\begin{equation}
\left( 1 -\frac{1}{3|w_{0}|}\right)\rho_{\rm crit}>(0.285\pm0.012)\rho_{\rm crit},
\end{equation}
or equivalently,
\begin{equation}
\left( 1 -\frac{1}{3|w_{0}|}\right)>(0.285\pm0.012).
\end{equation}
This condition holds as long as 
\begin{equation}\label{ineq}
|w_{0}|>\frac{1}{3-3((0.285\pm0.012))},
\end{equation}
which is not satisfied for the central values of the DESI fit (BAO plus CMB), 
since the right-hand side of the inequality is centered around $0.466$
(or $473$ if we assume $\Omega_m=\rho_0/\rho_{crit}=0.295$ and $0.5$
for $\Omega_m=0.334$).  This implies that $|w_0|$
must be greater than $0.466$ (or $0.473$, $0.5$) to sustain an accelerated
Universe at present time. This condition would be met by the DESI fit
without CMB data, where $|w_0|=0.55$. To account for the error bars, let us define the function
\begin{equation}
h(w_{0},\Omega_{m})=3w_{0}\Omega_{m}-(3w_{0}+1) >0,
\end{equation}
where we impose the condition for acceleration by demanding $h>0$. Using Gaussian error propagation, we obtain
\begin{equation}
h=-0.03475^{+0.7}_{-0.5}.
\end{equation}
\begin{figure}[ht!] 
    \includegraphics[width=0.4\textwidth]{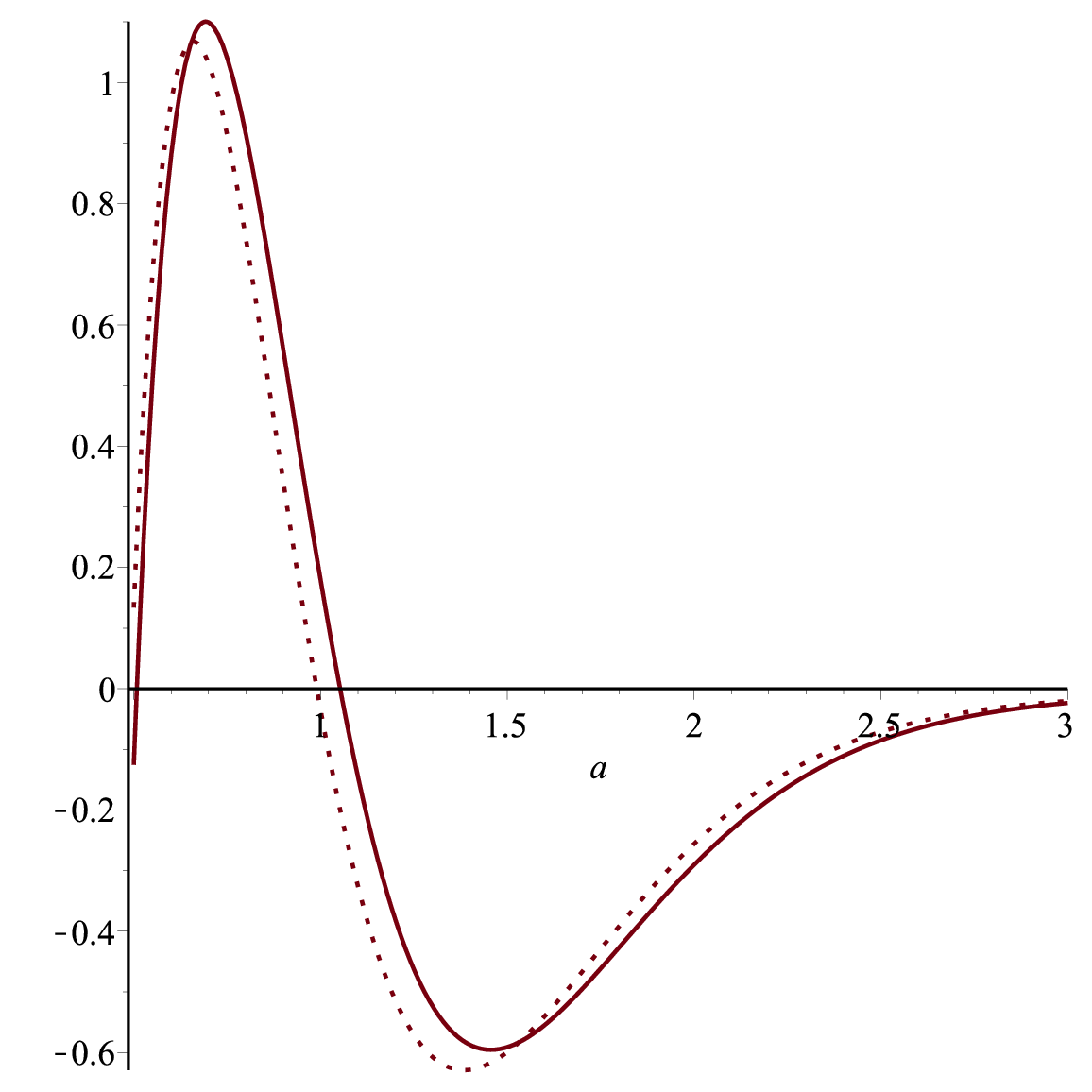}
    \includegraphics[width=0.4\textwidth]{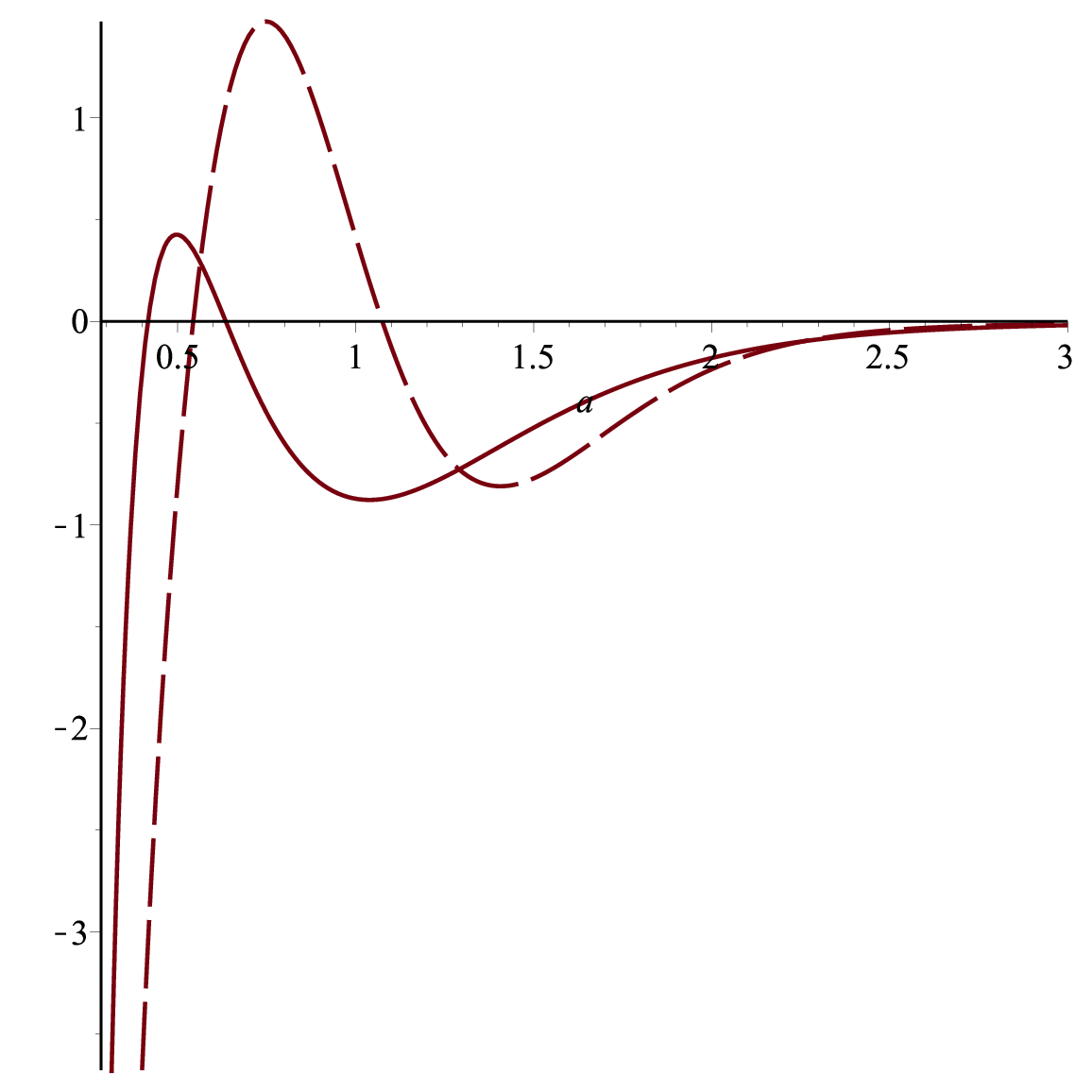}
\caption{\label{figure01}
Plots of \(\widetilde{q}=\frac{3\ddot{a}}{4\pi G a \rho_{crit}}\) as given by
equation \eqref{qttilde} with \(\gamma=1\) and \(P=0\) for different
values of \(w_0\) and \(w_a\). The parameter \(\rho_0\) in equation
\eqref{rho} is set to the central value of $0.285 \rho_{crit}$. The left panel displays the cases \(w_a = -1.79\), \(w_0 = -0.55\) (solid line) with maximum at $a=0.69250$ and minimum at $a=1.45583$; \(w_a = -1.79\), \(w_0 = -0.45\) (dotted line) with minimum at $a=0.65945$ and maximum at $a=1.38474$. In the right panel we considered the cases: \(w_a = -1.31\), \(w_0 = -0.16\) (solid line) with maximum at $0.53633$ and minimum at $a=1.14251$, \(w_a = -2.79\), \(w_0 = -0.76\) (long dashed line) with maximum at $a=0.77397$ and minimum at $a=1.44985$.}
\end{figure}

\begin{figure}[ht!] 
    \includegraphics[width=0.4\textwidth]{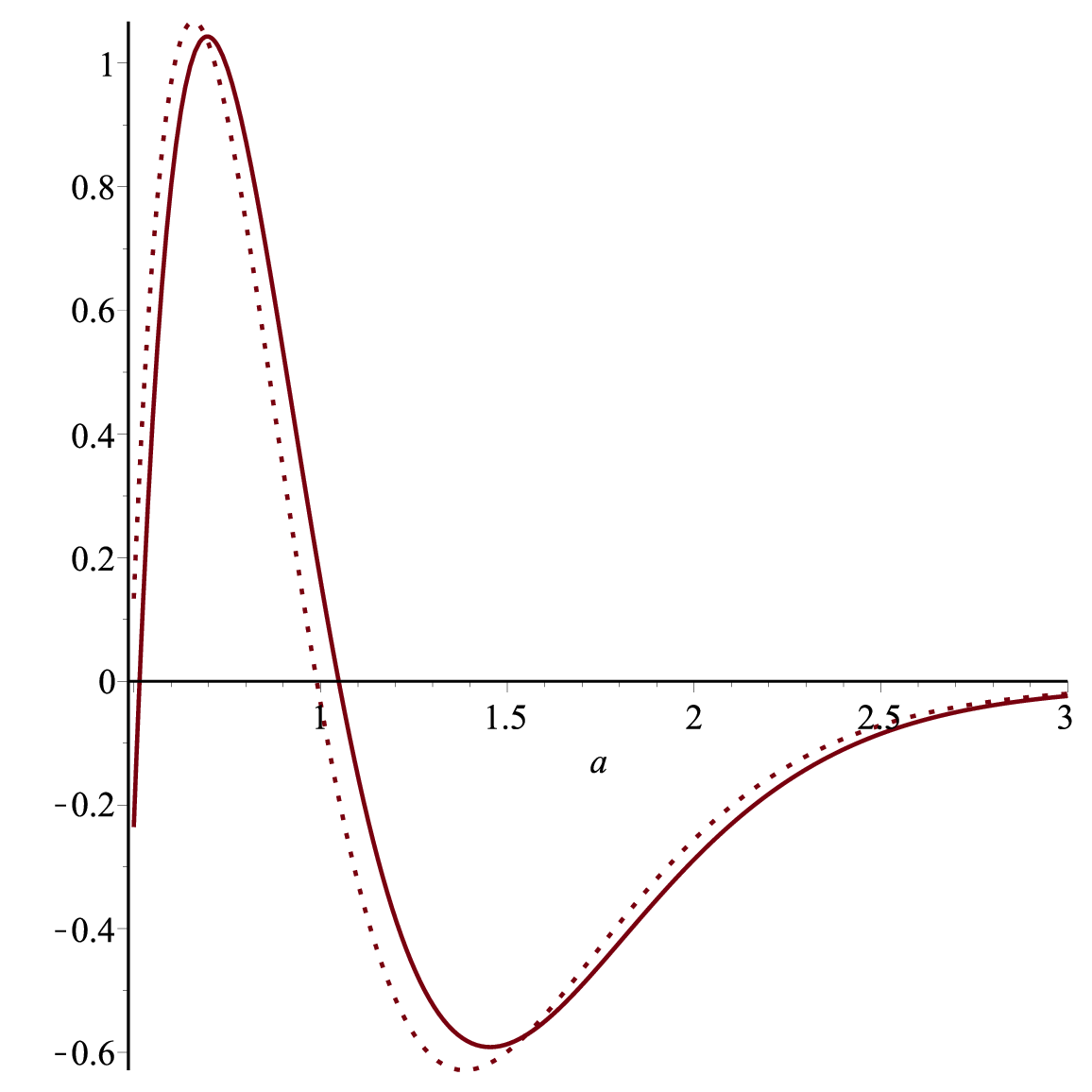}
    \includegraphics[width=0.4\textwidth]{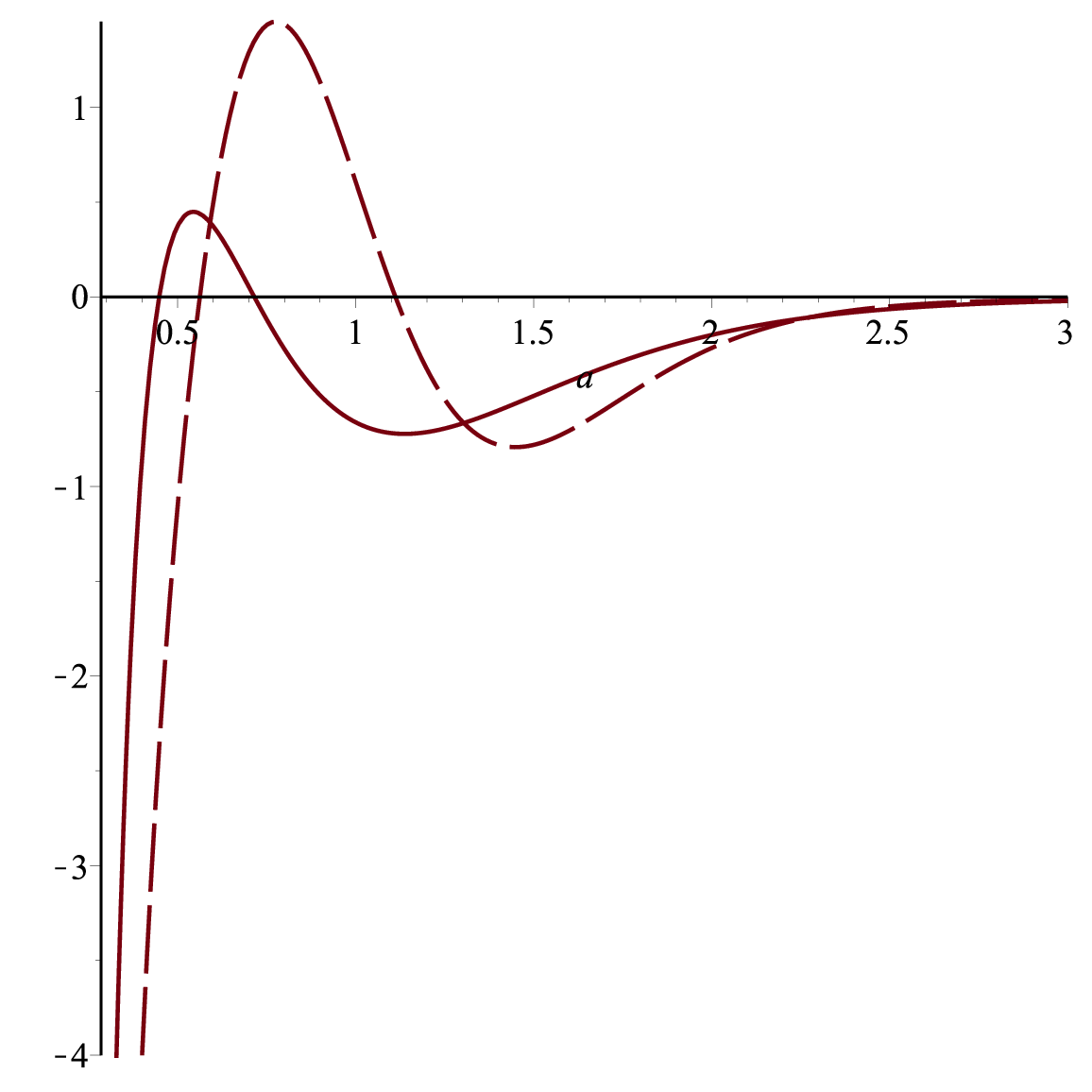}
\caption{\label{figure01a}
Plots of \(\widetilde{q}=\frac{3\ddot{a}}{4\pi G a \rho_{crit}}\) as given by
equation \eqref{qttilde} with \(\gamma=1\) and \(P=0\) for different
values of \(w_0\) and \(w_a\). The parameter \(\rho_0\) in equation
\eqref{rho} is set to the value of $0.295 \rho_{crit}$. The left panel displays the cases \(w_a = -1.79\), \(w_0 = -0.55\) (solid line) with maximum at $a=0.69691$ and minimum at $a=1.45387$; \(w_a = -1.79\), \(w_0 = -0.45\) (dotted line) with minimum at $a=0.66412$ and maximum at $a=1.38257$. In the right panel we considered the cases: \(w_a = -1.31\), \(w_0 = -0.16\) (solid line) with maximum at $0.54407$ and minimum at $a=1.13785$, \(w_a = -2.79\), \(w_0 = -0.76\) (long dashed line) with maximum at $a=0.77627$ and minimum at $a=1.44889$.}
\end{figure}

Starting from equation \eqref{F2D}, we can rewrite it using the equations of state as follows
\begin{equation}
\frac{\ddot{a}}{a}=-\frac{4\pi G}{3}\left[(3\gamma-2)\rho+(1+3w_{0}+3w_{a}(1-a))\rho_{\phi} \right].
\end{equation}
This allows us to redefine a new dimensionless quantity by expressing the densities in terms of $a$ as follows
\begin{eqnarray}
\widetilde{q}&\equiv& \frac{3\ddot{a}}{4\pi G a \rho_{\rm crit}},\\
&=&-\frac{1}{2}\left[(3\gamma-2)\Omega_{m}a^{-3\gamma}+(1+3w_{0}+3w_{a}(1-a))(1-\Omega_{m})a^{-3(1+w_{o}+w_{a})}e^{3w_{a}(a-1)} \right].\label{qttilde}
\end{eqnarray}
We note that $\ddot{a}>0$ (and thus $\tilde{q}>0$) corresponds to an accelerated stage. For
$\gamma=1$, we plot $\tilde{q}$ for different values of $w_{a}$,
$w_{0}$, and the central value of $\Omega_{m}$, and observe that at $a_{0}=1$ (the present stage), acceleration occurs only in certain cases (see Fig.~\ref{figure01} and Fig.~\ref{figure01a}).

\begin{figure}[ht!] 
    \includegraphics[width=0.45\textwidth]{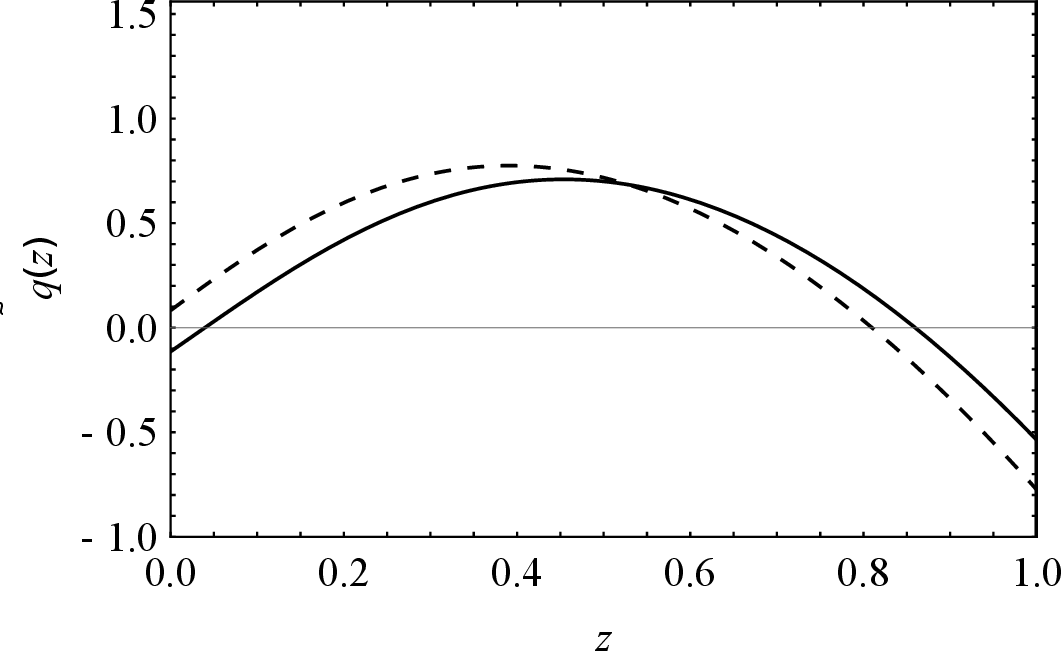}
    \includegraphics[width=0.45\textwidth]{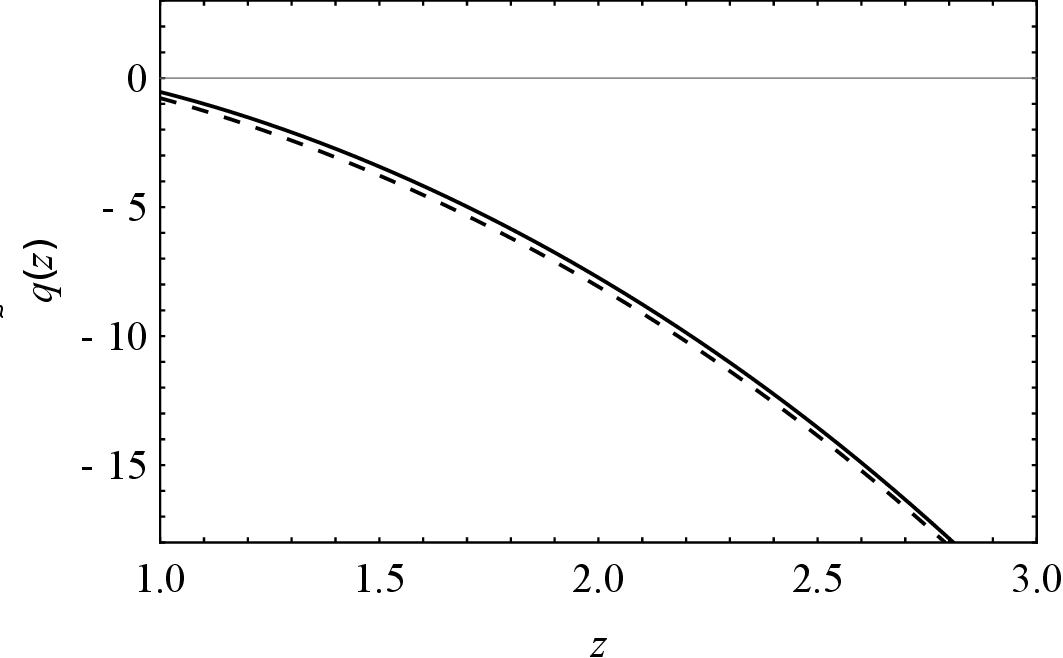}
\caption{\label{figure02}
Plots of $\tilde{q}$ in terms of redshift $z$ with \(\gamma=1\) and \(\Omega_{m}=0.334\). The left panel displays the cases \(w_a = -1.79\), \(w_0 = -0.55\) (dashed line); \(w_a = -1.79\), \(w_0 = -0.45\) (solid line) from $z=0$ to $z=1$. The right panel displays the same cases for $z$ from 1 to 3.}
\end{figure}

\begin{figure}[ht!] 
    \includegraphics[width=0.45\textwidth]{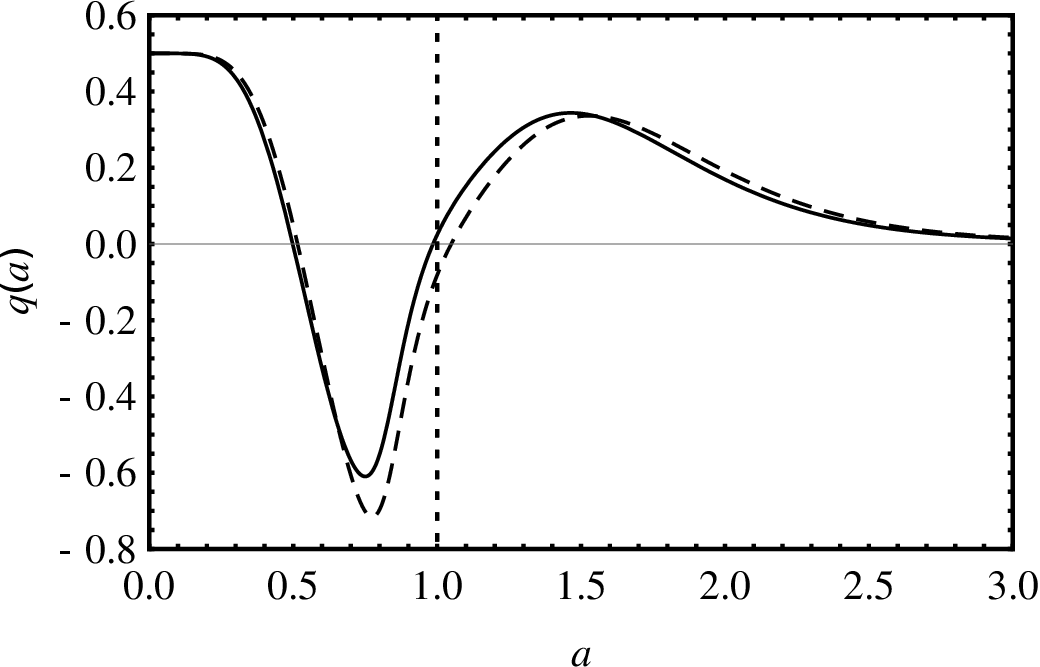}
    \includegraphics[width=0.45\textwidth]{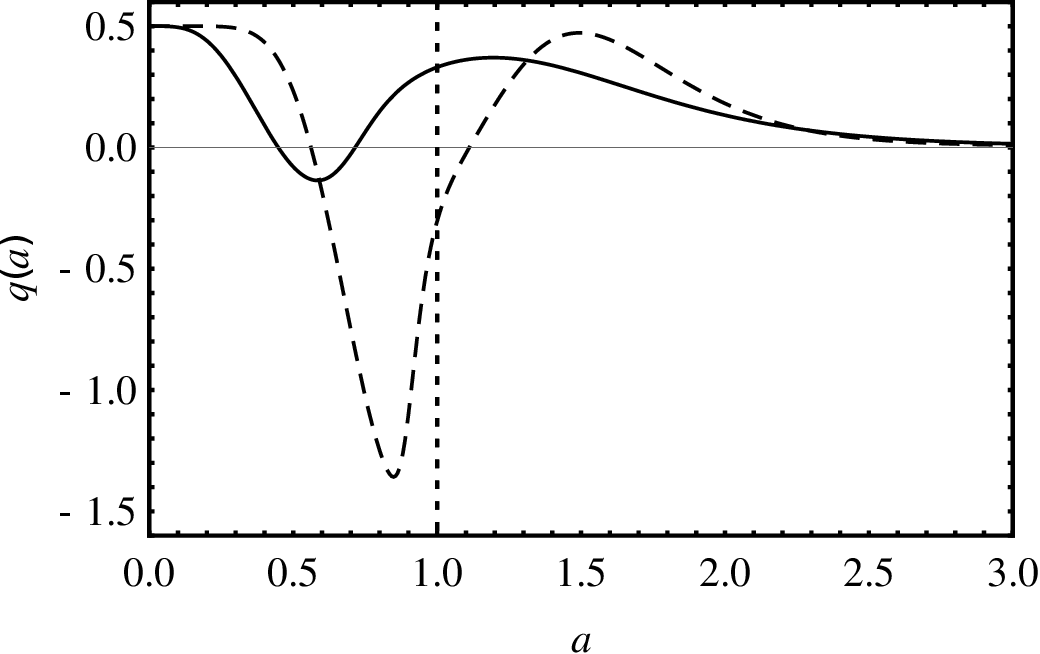}
\caption{\label{figure02bis}
Plots of $q$ as given by equation (\ref{qa}) with \(\gamma=1\) and \(\Omega_{m}=0.295\). The left panel displays the cases \(w_a = -1.79\), \(w_0 = -0.55\) (dashed line); \(w_a = -1.79\), \(w_0 = -0.45\) (solid line). In the right panel we considered the cases: \(w_a = -1.31\), \(w_0 = -0.16\) (solid line), \(w_a = -2.79\), \(w_0 = -0.76\) (dashed line).}
\end{figure}

An important parameter related to the acceleration of the Universe is the
deceleration parameter $q$ \cite{Visser}. This parameter is defined as a criterion to determine the accelerated stages of the Universe. In our case, $q$ is expressed as follows
\begin{eqnarray}
q&\equiv& -\frac{\ddot{a}a}{\dot{a}^{2}},\\
&=&\frac{1}{2}\left[1+\frac{3\Omega_{m}(\gamma-1)a^{-3\gamma}+3(1-\Omega_{m})(w_{0}+w_{a}(1-a))a^{-3(1+w_{o}+w_{a})}e^{3w_{a}(a-1)}}{\Omega_{m}a^{-3\gamma}+(1-\Omega_{m})a^{-3(1+w_{o}+w_{a})}e^{3w_{a}(a-1)}} \right].\label{qtilde}
\end{eqnarray}
A negative value of $q$ clearly indicates an accelerated stage. Therefore, it is useful to plot $q(a)$ for the case where $\gamma=1$, which is given by
\begin{equation}\label{qa}
q(a)_{\gamma=1}=\frac{1}{2}\left[1+3\frac{(1-\Omega_{m})(w_{0}+w_{a}(1-a))a^{-3(w_{0}+w_{a})}e^{3w_{a}(a-1)}}{\Omega_{m}+(1-\Omega_{m})a^{-3(w_{0}+w_{a})}e^{3w_{a}(a-1)}} \right].
\end{equation}
We observe that at $a_{0}=1$, the parameter is positive for the central values of $w_{0}$, $w_{a}$, and $\Omega_{m}$, suggesting that the Universe is not currently in an accelerated stage (see Fig.~\ref{figure02} and Fig.~\ref{figure02bis}).


Notice that $\tilde{q}(a)$ and $q(a)$ are different functions, but
qualitatively give the same information on the acceleration.
To explore the global behaviour across different scale factors $a$, we
have plotted $\tilde{q}=3\ddot{a}/(4\pi G \rho_{crit})$ as a function of $a$ in
Fig.~\ref{figure01} and Fig.~\ref{figure01a}. It is clear that at
$a=1$, the acceleration is not always positive. More importantly, the
behaviour of the acceleration stands in stark contrast to the
$\Lambda$CDM Universe.  Indeed, in terms of acceleration, the two Universes differ significantly from each other.

In particular, from Figs. 1 and 2 it is evident that the regions of positive
acceleration change while allowing different values 
permitted by the fit. Here a small comment about the choices of parameters is due.  According
to Fig.6 of reference \cite{DESI} the values of $w_0$ and $w_a$ are
correlated. This makes the choice of $w_a=-1.31$ and
$w_0=-0.16$ as well as $w_a=-2.79$ and $w_0=-0.76$ slightly outside the
contours, but it is done here on purpose to display how the
revelant physical quantities change if we move in this direction of the
parameter space. This comment refers also to the rest of the paper.

Whereas for the central values of the fit the
positive acceleration happens roughly in the region $0.5<a<
1$, choosing $w_a=-1.31$ and
$w_0=-0.16$ makes this range shorter.

An
important parameter in astrophysics is the cosmological redshift $z$
related to the scale factor by $a=(1+z)^{-1}$.  The behavior of
$\tilde{q}$ with $z$ is shown in Fig. 3.  To compare these results
with the corresponding ones of 
$\Lambda$CDM model we can readily infer that
\begin{equation} \label{qtildeLambda}
\tilde{q}_{\Lambda CDM}=-[\Omega_ma^{-3}-2\Omega_{\Lambda}]=2\Omega_{\Lambda}-\Omega-m(1+z)^3
\end{equation}
Using Planck data this tells us that in the $\Lambda$CDM model the
Universe is always accelerated provided $a>0.613$ ($z < 0.63$).  In
the Quntessence-DESI model the Universe is mostly accelerated in the
past roughly below $a<1$ and remains positive up to $z \simeq 0.85$,
i.e., up to a higher value of $z$ as compared to the $\Lambda$CDM
model. This fact could eventually be used to discriminate the models.

The parameter $q$ has been plotted in Fig. 4.
At $a=a_{0}=1$, the current value of $q$ has been determined in
\cite{Cosmographic}. Some of the values for this parameter are model-independent and we list them in Table~\ref{qtable}.  For the DESI Quintessence model, we can also calculate the current value of $q_{0}$, as follows
\begin{equation}
q(a=1)=\frac{1}{2}\left[ 1+3w_{0}(1-\Omega_{m})\right],
\end{equation}
which, for the central values $w_{0}=-0.45$ and $\Omega_{m}=0.295$, yields  $(q_{0})_{\rm DESI}=0.024125$, while for $\Omega_{m}=0.344$ it yields $(q_{0})_{\rm DESI}=0.0572$.

\begin{table}[ht!]
\begin{tabular}{|lllll|}
\hline
\multicolumn{5}{|c|}{Hubble data}                                                                                                      \\ \hline
\multicolumn{1}{|c|}{Model} & \multicolumn{1}{c|}{$H_{exp}$} & \multicolumn{1}{c|}{GP} & \multicolumn{1}{c|}{GA} & \multicolumn{1}{c|}{$\Lambda$CDM} \\ \hline
\multicolumn{1}{|l|}{$q_{0}$}     & \multicolumn{1}{l|}{-1.070$\pm$0.093}     & \multicolumn{1}{l|}{-0.856$\pm$0.111}   & \multicolumn{1}{l|}{-0.545$\pm$0.107}   & -0.645$\pm$0.023                         \\ \hline
\multicolumn{5}{|c|}{Pantheon data}                                                                                                    \\ \hline
\multicolumn{1}{|c|}{Model} & \multicolumn{1}{c|}{$H_{exp}$} & \multicolumn{1}{c|}{GP} & \multicolumn{1}{c|}{GA} & \multicolumn{1}{c|}{$\Lambda$CDM} \\ \hline
\multicolumn{1}{|l|}{$q_{0}$}     & \multicolumn{1}{l|}{-0.616$\pm$0.105}     & \multicolumn{1}{l|}{-0.558$\pm$0.040}   & \multicolumn{1}{l|}{-0.466$\pm$0.244}   &         -0.572$\pm$0.018                 \\ \hline
\end{tabular}
\caption{Values of the present deceleration parameter $q_{0}$ \cite{Cosmographic}.}
\label{qtable}
\end{table}
It is clear that the value of the deceleration parameter for DESI differs
from the corresponding values in Table~\ref{qtable}.

\subsection{Solution for the scale factor}
In this section, we will explore both numerical and analytical solutions for the scale factor.

\subsubsection{Numerical solution for the scale factor $a$}
\begin{figure}[ht!] 
    \includegraphics[width=0.4\textwidth]{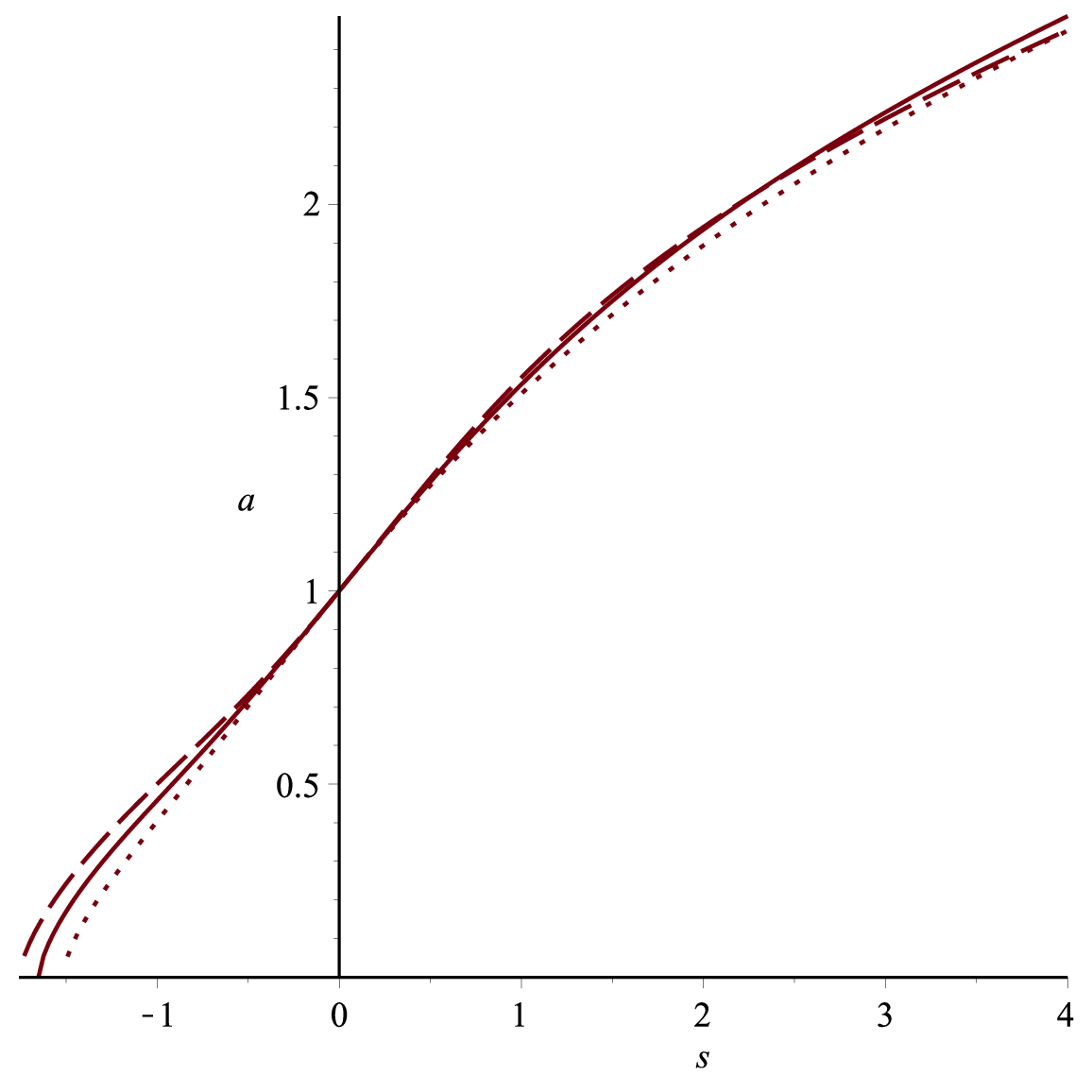}
\caption{\label{figure02a}
Plots of the scale factor $a$ as a function of the Hubble time $s$ for $\gamma=1$, $\Omega_m=0.285$ and different values of \(w_0\) and \(w_a\).  We considered the cases  \(w_a = -1.79\), \(w_0 = -0.55\) (solid line), \(w_a = -1.31\), \(w_0 = -0.16\) (dotted line), and \(w_a = -2.79\), \(w_0 = -0.76\) (dashed line).}
\end{figure}

\begin{figure}[ht!] 
    \includegraphics[width=0.4\textwidth]{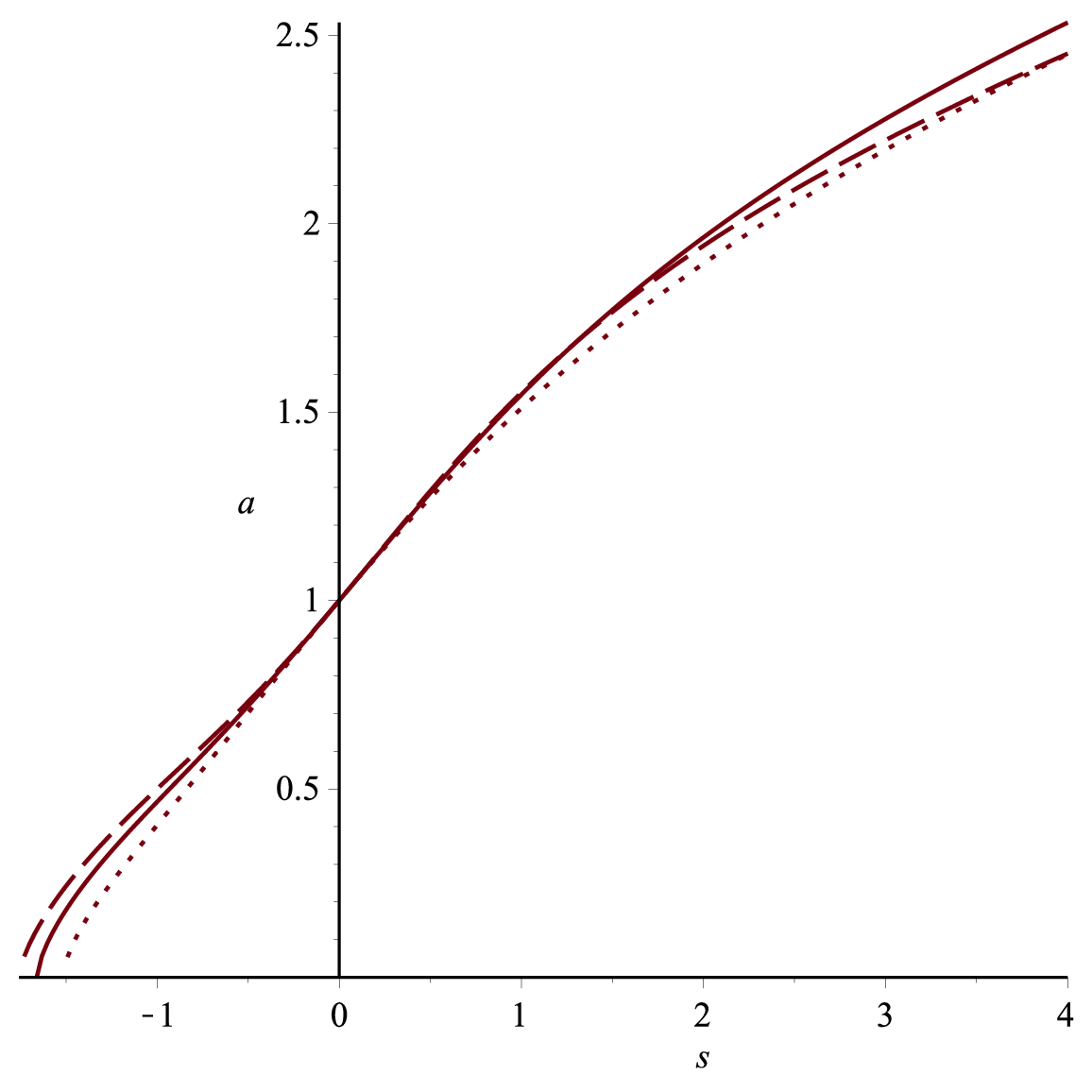}
\caption{\label{figure02b}
Plots of the scale factor $a$ as a function of the Hubble time $s$ for $\gamma=1$, $\Omega_m=0.295$ and different values of \(w_0\) and \(w_a\).  We considered the cases  \(w_a = -1.79\), \(w_0 = -0.55\) (solid line), \(w_a = -1.31\), \(w_0 = -0.16\) (dotted line), and \(w_a = -2.79\), \(w_0 = -0.76\) (dashed line).}
\end{figure}

By introducing the Hubble time $s=H_0 t$, we can rewrite \eqref{F1D} as 
\begin{equation}
    \left(\frac{a^{'}}{a}\right)=\frac{8\pi G}{3H_0^2}(\rho_\phi+\rho),
\end{equation}
where the prime denotes differentiation with respect to $s$. Moreover, considering that $\Omega_m=\rho_0/\rho_{crit}$ and $\rho_{\phi,0}=\rho_{crit}-\rho_0$, along with \eqref{rho} and \eqref{rhoPhi}, we arrive at the following initial value problem
\begin{equation}\label{glei}
    \frac{a^{'}}{a}=\sqrt{\frac{\Omega_m}{3a^3}+\frac{1-\Omega_m}{3}a^{-3(1+w_0+w_a)}e^{3w_a(a-1)}},\quad
    a(0)=1.
\end{equation}
We solved the equation above using Maple and the Runge-Kutta-Fehlberg
method \cite{Press}. Figure~\ref{figure02a} shows the behaviour of the scale factor for different parameter choices entering in equation \eqref{glei}. The solutions represented by the solid, dotted and dashed lines cannot be further computed beyond $a=-1.65199$, $a=-1.51944$, and $a=-1.75824$, respectively. The case $\omega_m=0.295$ is shown in Figure~\ref{figure02b}. The solutions corresponding to the solid, dotted and dashed lines cannot be further computed beyond $a=-1.66141$, $a=-1.50890$, and $a=-1.73886$, respectively.

\begin{figure}[ht!]
    \includegraphics[width=0.4\textwidth]{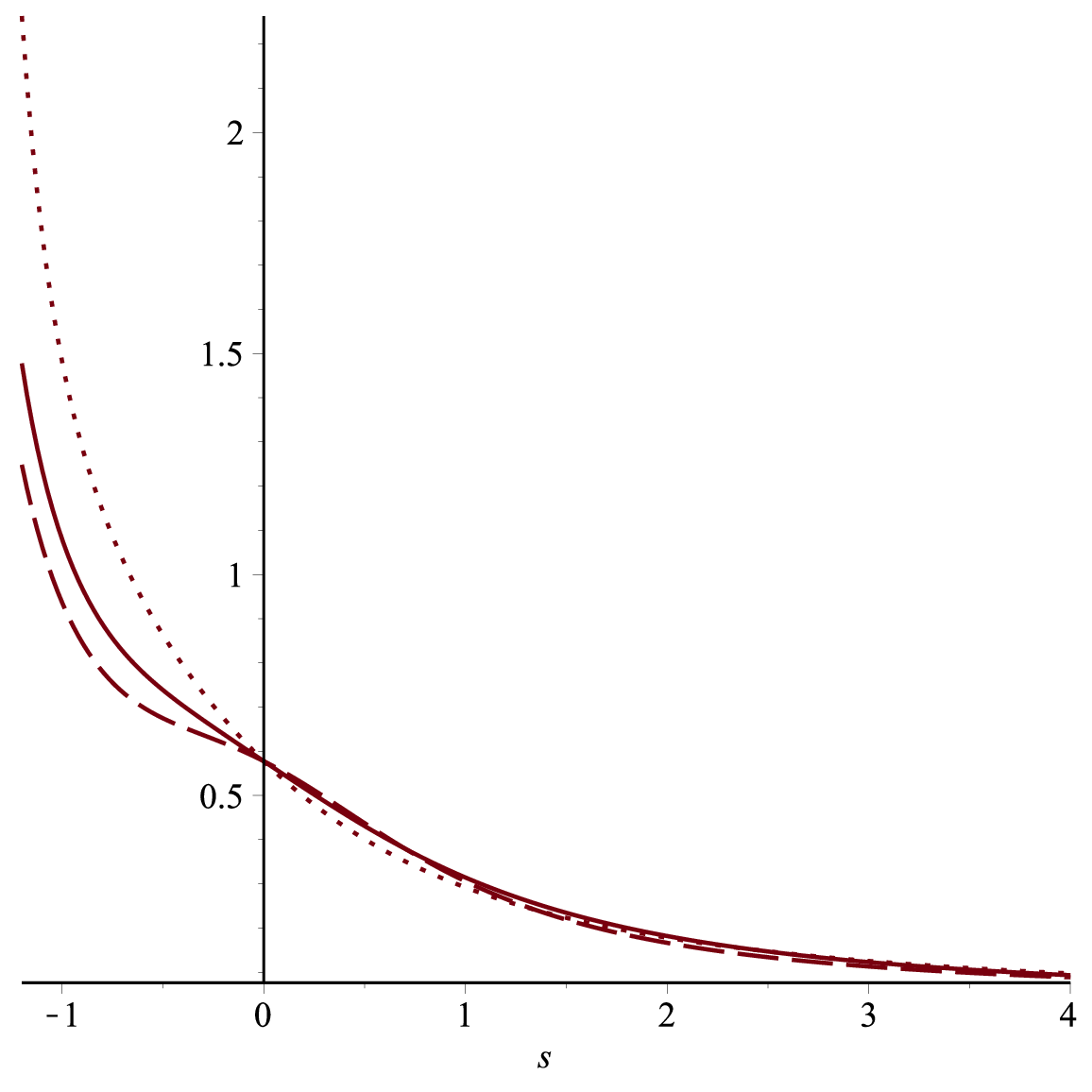}
\caption{\label{figureHubble2}
Plots of the Hubble parameter $H=a^{'}(s)/a(s)$   as a function of the Hubble time $s$ for $\gamma=1$, $\Omega_m=0.295$ and different values of \(w_0\) and \(w_a\).  We considered the cases  \(w_a = -1.79\), \(w_0 = -0.55\) (solid line), \(w_a = -1.31\), \(w_0 = -0.16\) (dotted line), and \(w_a = -2.79\), \(w_0 = -0.76\) (dashed line).}
\end{figure}

\begin{figure}[ht!] 
    \includegraphics[width=0.4\textwidth]{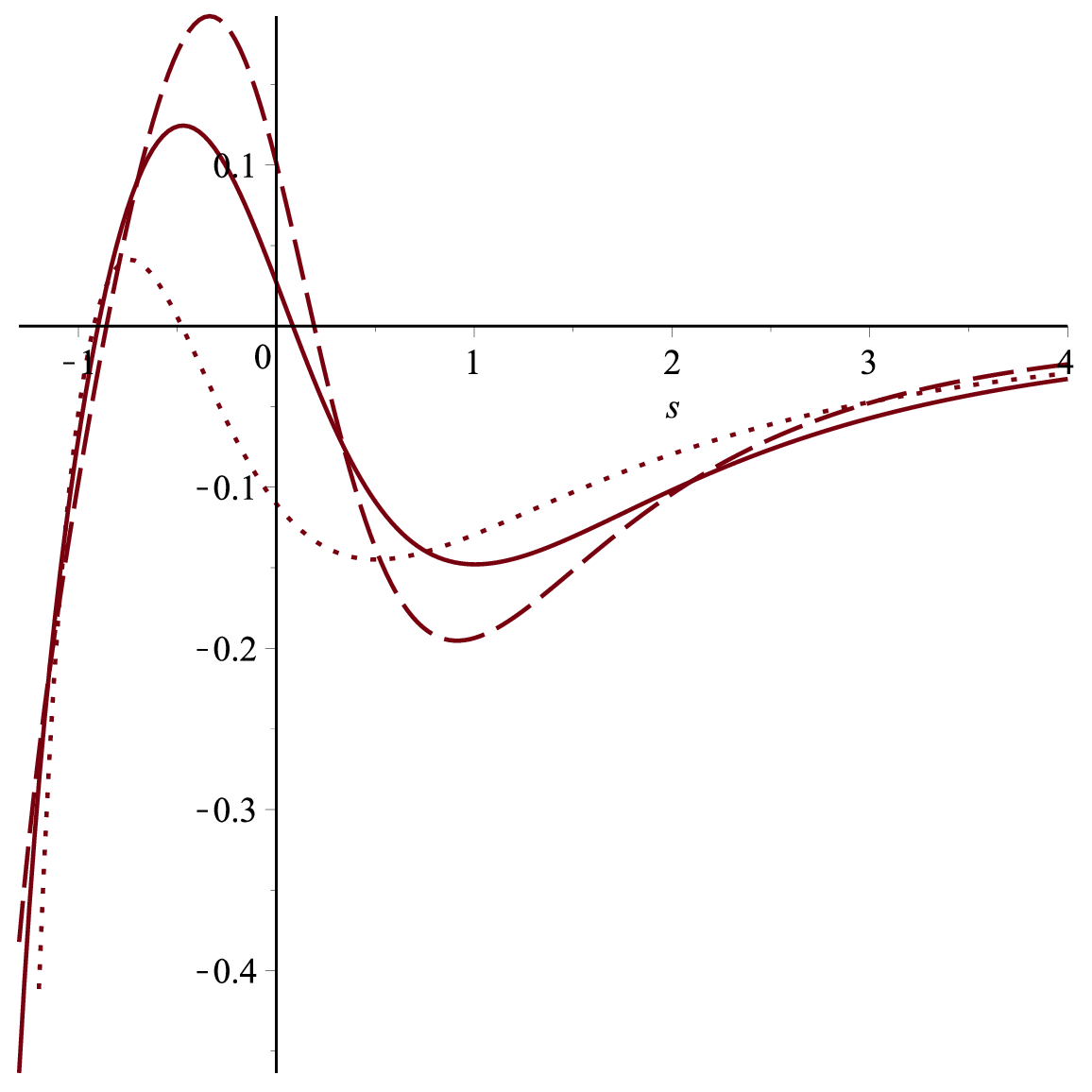}
\caption{\label{figure03b}
Plots of the acceleration $a^{''}$ as a function of the Hubble time $s$ for $\gamma=1$, $\Omega_m=0.295$ and different values of \(w_0\) and \(w_a\).  We considered the cases  \(w_a = -1.79\), \(w_0 = -0.55\) (solid line), \(w_a = -1.31\), \(w_0 = -0.16\) (dotted line), and \(w_a = -2.79\), \(w_0 = -0.76\) (dashed line).}
\end{figure}

We have plotted the solution for $\Omega_m=0.285$ (see Fig.~\ref{figure02a})  and $\Omega_m=0.295$ (see Fig.~\ref{figure02b}), even though the differences between the two cases are minimal. However, it is important to explicitly observe this fact. A careful inspection of these solutions reveals that they exhibit two inflection points, consistent with our discussion
on acceleration. The curve begins concave, then becomes convex, and finally turns concave again. This behaviour contrasts with the $\Lambda$CDM solution in
(\ref{solutiona2}), which is characterized by a shifted $sinh$ function with
a single inflection point. We take this opportunity to revisit the acceleration's independence from the Hubble time $s$. This is indeed confirmed, as shown in Fig.~\ref{figure03b}. Both this figure and Fig.~\ref{figureHubble2} were generated from the numerical solution for $a(s)$
using Maple18. In Fig.~\ref{figureHubble2}, we have plotted the
Hubble parameter $H$ versus the Hubble time $s$. Interestingly, for
different parameter choices in the DESI fit, $H$ remains insensitive for $S>0$, but shows significant differences in the range $s<0$. It is in this region where the behaviour diverges considerably from the solution (\ref{solutionH2}) of the cosmological constant model.

In the following, we will construct approximate analytical solutions for small and large values of $a$, as well as for $a$ around $1$.

\subsubsection{Solution for $a$ around 1}
Let $\gamma=1$. We first rewrite \eqref{equa} as follows
\begin{equation}\label{eq1}
    \frac{da}{dt}=\pm\sqrt{\frac{A+Ba^\alpha e^{\beta(a-1)}}{a}}   
\end{equation}
where
\begin{equation}
     A=\frac{8}{3}\pi G\rho_0,\quad
    B=\frac{8}{3}\pi G\rho_{\phi,0},\quad
    \alpha=-3(w_0+w_a)>0,\quad
    \beta=3w_a<0,
\end{equation}
and with the initial condition $a(0)=1$. Using Maple, we obtain the following series solution in a neighbourhood of $a(0)=1$ 
\begin{eqnarray}
    a(t)&=&1\pm\sqrt{A+B}t-\frac{1}{4}\left[A+(1+3w_0)B\right]t^2\nonumber\\
    &&\pm\frac{\sqrt{A+B}}{12}\left[2(A+B)+3B(3w_0^2+3w_0+w_a)\right]t^3+\mathcal{O}(t^4).
\end{eqnarray}
Differentiating the above expression twice with respect to the time variable, we find that around $a(0)=1$ the acceleration behaves as 
\begin{equation}
    \ddot{a}(t)=-\frac{1}{2}\left[A+(1+3w_0)B\right]\pm\frac{\sqrt{A+B}}{2}\left[2(A+B)+3B(3w_0^2+3w_0+w_a)\right]t+\mathcal{O}(t^2).
\end{equation}
Note that the requirement $\ddot{a}(0)>0$ implies that the term $A+(1+3w_0)B$ must be negative. It is straightforward to check that this inequality is equivalent to \eqref{ineq}.

\subsubsection{Solution for large $a$}
Let $\gamma=1$ and rewrite the argument of the square root in \eqref{eq1} as follows
\begin{eqnarray}
   \frac{A+Ba^\alpha e^{\beta(a-1)}}{a}&\approx&\frac{A}{a}+Ba^{\alpha-1}e^{-|\beta|a},\\
   &=&\frac{8}{3}\pi G\left(\frac{\rho_0}{a}+\rho_{\phi,0}a^{\alpha-1}e^{-|\beta|a}\right).
\end{eqnarray}
According to the CMB data, the term $\alpha-1$ is positive and can take on values in the range $[3.11,9.35]$. Furthermore, $\beta$ is negative,  indicating an exponential decay as $a$ increases. Substituting  $\rho_0=\Omega_m\rho_{crit}$ and $\rho_{\phi,0}=(1-\Omega_m)\rho_{crit}$ gives 
\begin{equation}
    \frac{A+Ba^\alpha e^{\beta(a-1)}}{a}\approx\frac{8}{3a}\pi G\rho_{crit}\left[\Omega_m+(1-\Omega_m)a^{\alpha}e^{-|\beta|a}\right].
\end{equation}
Thus, in the regime where $a\gg 1$, the right-hand side of equation \eqref{equa} can be expressed as
\begin{equation}
    \frac{A+Ba^\alpha e^{\beta(a-1)}}{a}\approx\frac{8}{3a}\pi G\rho_{crit}\Omega_m.
\end{equation}
As a result, $a$ asymptotically satisfies the following equation
\begin{equation}
    \frac{da_\infty}{dt}=\pm\sqrt{\frac{K}{a_\infty}},\quad K=\frac{8}{3}\pi G\rho_{crit}\Omega_m.
\end{equation}
Integrating this equation yields
\begin{equation}
    \frac{2}{3}a_\infty^{3/2}=\pm\sqrt{K}t+C,
\end{equation}
where $C$ is an arbitrary integration constant that can be neglected when $t\gg 1$. This leads to the asymptotic solution
\begin{equation}
    a_\infty(t)\approx\sqrt[3]{12\pi G\rho_{crit}\Omega_m}t^{2/3}.
\end{equation}

\begin{figure}[ht!] 
    \includegraphics[width=0.4\textwidth]{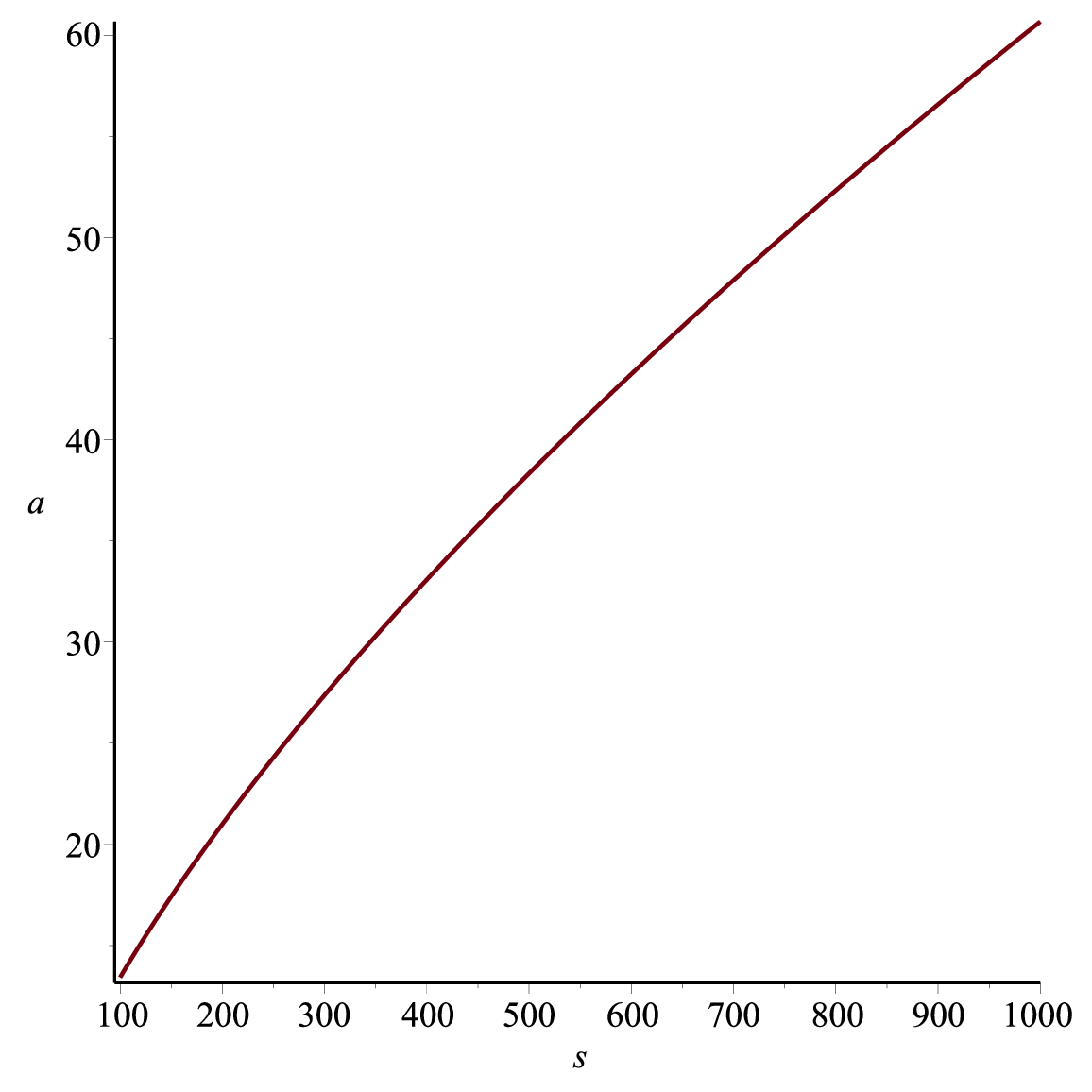}
\caption{\label{figure04b}
Plots of the scale factor $a$ as a function of the Hubble time $s$ for $\gamma=1$, $\Omega_m=0.295$, \(w_a = -1.79\), and \(w_0 = -0.45\).}
\end{figure}

As Dark Energy diminishes and eventually vanishes at large times, the model transitions to a simple cosmology with $k=0$ and $\Lambda=0$, preventing any collapse \cite{Nar}. This behaviour is also illustrated in Fig.~\ref{figure04b}.

\subsubsection{Solution for small $a$}
In the scenario of small $a$, we can assume we are in the radiation epoch, where  $\gamma=4/3$. First, we rewrite \eqref{equa} as follows
\begin{equation}\label{eq2}
    \frac{da}{dt}=\pm\frac{\sqrt{A+Ba^{1+\alpha}e^{\beta(a-1)}}}{a}
\end{equation}
with $A$, $B$, $\alpha$ and $\beta$ defined as in \eqref{eq1}. Given that $1+\alpha$ is positive, as confirmed by the CMB data, we can Taylor expand the numerator in \eqref{eq2}, leading to the following simplified differential equation
\begin{equation}
    \frac{da_0}{dt}=\pm\frac{\sqrt{A}}{a_0}.
\end{equation}
The corresponding solution is straightforward to obtain and is given by
\begin{equation}
    a_0(t)=\sqrt{2\left(C\pm\sqrt{\frac{8}{3}\pi G\rho_{crit}\Omega_m}t\right)},
\end{equation}
where $C$ is an integration constant chosen such that $a(T)=0$ for some $T<0$.

\subsection{Lifetime of the Universe}\label{4D}
The bounds on the lifetime of the Universe serves as a powerful tool for evaluating the validity of cosmological models \cite{our}, as the Universe cannot be younger than the objects it contains. Of particular relevance are the lower bounds on the Universe's age, especially when considering some of the oldest galaxies and stars detected in our observable Universe. For the oldest galaxies, the lower limit is approximately $12.5$ Gyr \cite{Age1,Age2}, while some of the oldest stars have been estimated to be even older, with a lower limit ranging between $13.2$ and $13.53$ Gyr \cite{Age3,Age4,Age5}. These ages have been determined using chemodynamical and Population III models, and they appear to be influenced by data from the CMB. 
 
In the DESI Quintessence model, the lifetime of the Universe can be determined by
\begin{equation}
T_{\rm Univ}=\left|\int_{a=a_{0}=1}^{a=0}\frac{da}{a\sqrt{f(a)}}\right|.
\end{equation}
Considering that $\rho_{crit}=3H_0^2/8\pi G$, and introducing the dimensionless quantity $\Omega_m=\rho_0/\rho_{crit}$ with  $\rho_{\phi,0}=\rho_{crit}-\rho_0$, we can rewrite the integral as 
\begin{equation}
T_{Univ}H_0=\left|\int_0^1\frac{da}{a\sqrt{\Omega_m a^{-3}+(1-\Omega_m)a^{-3(1+w_0+w_a)}e^{3w_a(a-1)}}}\right|.
\end{equation}
The numerical values corresponding  to this equation, using different
allowed values for $w_{a}$ and $w_{0}$ are presented in the Tables
V-VIII. Table V corresponds  to the actual DESI fit with variable
equation of state. The other tables show the differences when we
change the Hubble constant and/or the matter density.  

\begin{table}[ht]
\caption{Lifetime of the Universe in DESI Quintessence model with $\Omega_m=0.344$}
\begin{tabular}{|cc|cc|}
\hline
\multicolumn{2}{|c|}{$H_{0} = (64.7^{+2.2}_{-3.3})$ Km s$^{-1}$ Mpc$^{-1}$}       & \multicolumn{2}{c|}{$\frac{1}{H_{0}}$=(15.11$^{+0.8}_{-1.2}$) Gyr}       \\ \hline
\multicolumn{1}{|c|}{$w_{a}$}  & $w_{0}$  & \multicolumn{1}{c|}{$T_{\rm Univ} \cdot H_{0}$}  & $T_{\rm Univ}$ (Gyr)  \\ \hline
\multicolumn{1}{|c|}{-1.79} & -0.45 & \multicolumn{1}{c|}{0.90652} & 13.70 \\ \hline
\multicolumn{1}{|c|}{-1.79-1.0} & -0.45-0.21 & \multicolumn{1}{c|}{0.9471} & 14.31 \\ \hline
\multicolumn{1}{|c|}{-1.79+0.48} & -0.45+0.34 & \multicolumn{1}{c|}{0.8346} & 12.61 \\ \hline
\end{tabular}
\label{LifetimeTableD}
\end{table}

The age of the oldest stars of $13.53$ Gyr comes very close to the
$13.7$ Gyr estimated in Table \ref{LifetimeTableD}. It all depends if
stars could have formed $170$ million years after the Big Bang.  We
also notice that the Hubble parameter corresponding to the fit used in
the Table \ref{LifetimeTableD} has a relative big error bar.  This
means that some allowed values of Hubble parameter will lead to a
smaller lifetime. For instance, using $H_0^{-1}=(15.11-0.6)$ Gyr we arrive
at lifetime of the universe of $13.15$ Gyr which is again smaller than
the age of the oldest stars.

\begin{table}[ht]
\caption{Lifetime of the Universe in DESI Quintessence model with $\Omega_m=0.285$}
\begin{tabular}{|cc|cc|}
\hline
\multicolumn{2}{|c|}{$H_{0} = (68.3\pm 1.1)$ km s$^{-1}$ Mpc$^{-1}$}       & \multicolumn{2}{c|}{$\frac{1}{H_{0}}$=(14.316 $\pm$ 0.2) Gyr}       \\ \hline
\multicolumn{1}{|c|}{$w_{a}$}  & $w_{0}$  & \multicolumn{1}{c|}{$T_{\rm Univ} \cdot H_{0}$}  & $T_{\rm Univ}$ (Gyr)  \\ \hline
\multicolumn{1}{|c|}{-1.79} & -0.45 & \multicolumn{1}{c|}{0.95378} & 13.65 \\ \hline
\multicolumn{1}{|c|}{-1.79-1} & -0.45-0.21 & \multicolumn{1}{c|}{1.00498} & 14.39 \\ \hline
\multicolumn{1}{|c|}{-1.79+0.48} & -0.45+0.34 & \multicolumn{1}{c|}{0.86531} & 12.39 \\ \hline
\end{tabular}
\label{LifetimeTable}
\end{table}

\begin{table}[ht]
\caption{Lifetime of the Universe in DESI Quintessence model with $\Omega_m=0.295$}
\begin{tabular}{|cc|cc|}
\hline
\multicolumn{2}{|c|}{$H_{0} = (68.3\pm 1.1)$ Km s$^{-1}$ Mpc$^{-1}$}       & \multicolumn{2}{c|}{$\frac{1}{H_{0}}$=(14.316 $\pm$ 0.2) Gyr}       \\ \hline
\multicolumn{1}{|c|}{$w_{a}$}  & $w_{0}$  & \multicolumn{1}{c|}{$T_{\rm Univ} \cdot H_{0}$}  & $T_{\rm Univ}$ (Gyr)  \\ \hline
\multicolumn{1}{|c|}{-1.79} & -0.45 & \multicolumn{1}{c|}{0.94499} & 13.53 \\ \hline
\multicolumn{1}{|c|}{-1.79-1.0} & -0.45-0.21 & \multicolumn{1}{c|}{0.99417} & 14.23 \\ \hline
\multicolumn{1}{|c|}{-1.79+0.48} & -0.45+0.34 & \multicolumn{1}{c|}{0.85966} & 12.31 \\ \hline
\end{tabular}
\label{LifetimeTableB}
\end{table}

\begin{table}[ht]
\caption{Lifetime of the Universe in DESI Quintessence model with $\Omega_m=0.344$}
\begin{tabular}{|cc|cc|}
\hline
\multicolumn{2}{|c|}{$H_{0} = (68.3\pm 1.1)$ Km s$^{-1}$ Mpc$^{-1}$}       & \multicolumn{2}{c|}{$\frac{1}{H_{0}}$=(14.316 $\pm$ 0.2) Gyr}       \\ \hline
\multicolumn{1}{|c|}{$w_{a}$}  & $w_{0}$  & \multicolumn{1}{c|}{$T_{\rm Univ} \cdot H_{0}$}  & $T_{\rm Univ}$ (Gyr)  \\ \hline
\multicolumn{1}{|c|}{-1.79} & -0.45 & \multicolumn{1}{c|}{0.90652} & 12.98 \\ \hline
\multicolumn{1}{|c|}{-1.79-1.0} & -0.45-0.21 & \multicolumn{1}{c|}{0.9471} & 13.56 \\ \hline
\multicolumn{1}{|c|}{-1.79+0.48} & -0.45+0.34 & \multicolumn{1}{c|}{0.8346} & 11.95 \\ \hline
\end{tabular}
\label{LifetimeTableC}
\end{table}
 However, it is also possible to consider the reverse conclusion. If
 the lifetime of a new cosmological model is smaller than the lifetime
 of the oldest stars we could equally conclude that such estimates of
the ages of the oldest stars may not be accurate as the latter are
model dependent. A more model independent approach is presented
below.

Returning to the
tables in Section~\ref{Sec2}, the Tables~\ref{LifetimeTableDESI1} and \ref{LifetimeTableDESI} present the results for the lifetime of the
universe for one of the DESI fits with a constant parameter $w$ in the
equation of state. Notably, some of the calculated
lifetimes (especially those in Table \ref{LifetimeTableDESI1}) within
the $\Lambda$CDM model face some challenges if we accept the current
estimates for the ages of the oldest stars.

Galaxies observed at a high cosmological redshift $z$ \cite{Age1, CosmicDawn} offer also a
good test of the validity of a given model. In this case the time needed to reach the
observer $\Delta t$ should always be smaller then the lifetime of the
universe.  One needs a relation between the cosmological redshift $z$
and $\Delta t$ which is readily obtained from
\begin{equation}
a=\frac{1}{1+z},
\end{equation}
and the definition of the Hubble function
\begin{equation}
H=\frac{1}{a}\frac{da}{dt}=(1+z)\frac{da}{dz}\frac{dz}{dt}
\end{equation}
This allows us to write
\begin{equation}
dt=-\frac{1}{H_{0}}\frac{dz}{(1+z)E(z)^{1/2}}
\end{equation}
where we used
\begin{equation}
\left(\frac{H(z)}{H_{0}}\right)^{2}\equiv E(z)
\end{equation}
Here $H(z)$ follows from the Friedmann equations. For instance, in the
$\Lambda$CDM model we have
\begin{equation}
\Delta t=\frac{1}{H_{0}}\int_{0}^{z_{e}} \frac{dz}{(1+z)\sqrt{\Omega_{m}(z+1)^{3}+\Omega_{\Lambda}}}=\frac{1}{H_{0}}\left[-\frac{2}{3\sqrt{\Omega_{\Lambda}}}\coth^{-1}\left(\sqrt{\frac{\Omega_{m}(1+z)^{3}+\Omega_{\Lambda}}{\Omega_{\Lambda}}}\right) \right] \Bigg|_{z=0}^{z=z_{e}}
\end{equation}
whereas the Quintessence model would give rise to
\begin{equation}
\Delta t=\frac{1}{H_{0}}\int_{0}^{z_{e}}\frac{dz}{(1+z)\left[ (1-\Omega_{m})(1+z)^{3(1+w_{0}+w_{a})}e^{-3w_{a}\frac{z}{z+1}}+\Omega_{m}(z+1)^{3}\right]^{1/2}}
\end{equation}
The lifetime of the universe is obtained in both formulas by taking
$z_e \to \infty$. The values of $\Delta t$ for several relevant
redshifts $z$ in both $\Lambda$CDM and Quintessence models are given
in Tables~\ref{RedshiftLambda} and \ref{RedshiftDESI}.  The evidence
of the luminous objects at very high redshift we have taken from
\cite{Laursen, Yan}. For $z=16-20$ we have used Fig.6 in
\cite{Laursen}. The value $z=24.7$ is mentioned in \cite{Yan}, but as
stated there is subject to the interpretation of the templates. At the
same time, this very high redshift has prompted the author to point
out that the next challenge is to find objects beyond $z=20$. The
values of the time span $\Delta t$ needed to reach us are displayed in
tables IX. 
and X. In general, there is a tension between the lifetime of the
Universe. If we change the values of $w_a$ and $w_0$ and push them to
the border of the allowed parameter space it is possible to obtain a
higher lifetime in the Quintessence-DESI model (see table V).  In both
models we face the problem of the so-called ``impossible early
galaxy'' formation \cite{Steinhardt1}. This could be due to the
cosmological model (a too short lifetime) or due to the interpretation
of the images \cite{Steinhardt2}. A third possibility is some unknown
physics at the early epoch of the Universe.
\begin{table}[]
\begin{scriptsize}
\caption{Time needed to reach an observer $\Delta t$ for high redshifts in different $\Lambda$CDM fits.}
\label{RedshiftLambda}
\begin{tabular}{|c|c|c|c|c|c|c|}
\hline
$\Lambda$CDM Fit                       & $H_{0}$ (Km s$^{-1}$ Mpc$^{-1}$)                & $\Omega_{m}$                & $\Omega_{\Lambda}$            & $z_{e}$ & $\Delta t H_{0}$ & $\Delta t$ (Gyrs)  \\ \hline
\multirow{4}{*}{Planck}   & \multirow{4}{*}{67.36 $\pm$ 0.54 } & \multirow{4}{*}{ \quad 0.3153 \quad} & \multirow{4}{*}{\quad 0.6847 \quad} & 12  & 0.9254   & 13.43  \\ \cline{5-7} 
                          &                   &                   &                   & $\quad 13 \quad$  &   0.9281   &  13.47 \\ \cline{5-7} 
                          &                   &                   &                   &  14 &  0.9303  & 13.50  \\ \cline{5-7} 
                          &                   &                   &                   &        $16$       &  0.9338  & 13.55  \\ \cline{5-7} 
                          &                   &                   &                   &        $20$       &  0.9384  & 13.62  \\ \cline{5-7} 
                          &                   &                   &                   &        $24.7$       &  0.9416  & 13.67  \\ \cline{5-7} 
                          &                   &                   &                   &        $\infty$       &  0.95073  & 13.80  \\ \hline
\multirow{4}{*}{DESI}     & \multirow{4}{*}{67.97 $\pm$ 0.38} & \multirow{4}{*}{0.3069} & \multirow{4}{*}{0.6931} & 12  & 0.9323   & 13.41  \\ \cline{5-7} 
                          &                   &                   &                   & 13  &   0.935 & 13.45  \\ \cline{5-7} 
                          &                   &                   &                   &  14 & 0.9373   &  13.48 \\ \cline{5-7} 
                          &                   &                   &                   &        $16$       &  0.9408  & 13.53  \\ \cline{5-7} 
                          &                   &                   &                   &        $20$       &  0.9454  & 13.60  \\ \cline{5-7} 
                          &                   &                   &                   &        $24.7$       &  0.9487  & 13.65  \\ \cline{5-7} 
                          &                   &                   &                   &        $\infty$       &  0.958  & 13.78  \\ \hline
\multirow{4}{*}{DESI+CMB+BAO} & \multirow{4}{*}{68.3 $\pm$ 1.1} & \multirow{4}{*}{0.349} & \multirow{4}{*}{0.651} & 12  & 0.8996   &  12.88 \\ \cline{5-7} 
                          &                   &                   &                   & 13  &  0.9022  & 12.92  \\ \cline{5-7} 
                          &                   &                   &                   &14    & 0.9043   & 12.95  \\ \cline{5-7} 
                          &                   &                   &                   &  $16$ & 0.9076   & 12.99  \\ \cline{5-7} 
                          &                   &                   &                   &  $20$ & 0.9119   & 13.06  \\ \cline{5-7} 
                          &                   &                   &                   &  $24.7$ & 0.91504   & 13.10  \\   \cline{5-7} 
                          &                   &                   &                   &  $\infty$ & 0.9237   & 13.22  \\ \hline
\end{tabular}
\end{scriptsize}
\end{table}

\begin{table}[]
\caption{Time needed to reach an observer $\Delta t$ for high redshifts in different non-constant equation of state fits ($w_{0}w_{a}\Lambda$CDM).}
\label{RedshiftDESI}
\begin{tabular}{|c|c|c|c|c|c|c|}
\hline
$\quad w_{0}w_{a}$$\Lambda$CDM Fit  $\quad$                           & $H_{0}$ (Km s$^{-1}$ Mpc$^{-1}$)               & $\quad \Omega_{m} \quad$                 & $\quad \Omega_{\Lambda} \quad$            &  $z_{e}$ & $\Delta t H_{0}$ & $\Delta t$ (Gyrs) \\ \hline
\multirow{4}{*}{DESI}           & \multirow{4}{*}{68.3 $\pm$ 1.1} & \multirow{4}{*}{0.344} & \multirow{4}{*}{0.656} & 12  & 0.8823   &  12.63 \\ \cline{5-7} 
                                    &                   &                   &                   & $\quad 13 \quad$  &   0.8848 & 12.67  \\ \cline{5-7} 
                                    &                   &                   &                   &  14 &  0.8870  & 12.70  \\ \cline{5-7} 
                                    &                   &                   &                   & $16$  & 0.8903   & 12.75  \\ \cline{5-7} 
                                    &                   &                   &                   & $20$  & 0.8947   & 12.81  \\ \cline{5-7} 
                                    &                   &                   &                   & $24.7$  & 0.8978   & 12.85  \\ \cline{5-7} 
                                    &                   &                   &                   & $\infty$  & 0.9065   & 12.98  \\ \hline
\multirow{4}{*}{DESI + CMB } & \multirow{4}{*}{64.7 $^{+2.2}_{-3.3}$ } & \multirow{4}{*}{0.344} & \multirow{4}{*}{0.656} & 12   & 0.8823    & 13.33   \\ \cline{5-7} 
                                    &                   &                   &                   & 13  &  0.8848  &  13.37  \\ \cline{5-7} 
                                    &                   &                   &                   & 14  & 0.8870   &  13.40 \\  \cline{5-7} 
                                    &                   &                   &                   & $16$ &0.8903    &  13.46 \\  \cline{5-7} 
                                    &                   &                   &                   & $20$ &0.8947    &  13.52 \\  \cline{5-7} 
                                    &                   &                   &                   & $24.7$ &0.8978    &  13.57 \\  \cline{5-7} 
                                    &                   &                   &                   & $\infty$ &0.9065    &  13.70 \\ \hline
\end{tabular}
\end{table}
 In passing, we note that the so-called "early galaxy problem" has prompted several authors to either modify the standard cosmological model \cite{Gupta1, Gupta2, Gupta3} or reconsider the interpretation of cosmological redshift \cite{Shamir}. In \cite{Gupta1}, a novel cosmological model is proposed in which fundamental constants, specifically Newton's gravitational constant \(G_N(t)\) and the speed of light \(c(t)\), are allowed to vary with time. This modification of the Friedmann equations yields a significantly older Universe, with an age of $26.7$ Gyr, and suggests that at redshifts \(z=10\) and \(z=20\), the Universe would already be $5.8$ Gyr and $3.5$ Gyr old, respectively, and therefore, it would potentially provide enough time for galaxy formation and evolution. While these results are promising from the perspective of gravitational cosmology, introducing time-varying constants also impacts astroparticle physics, particularly early-universe processes such as nucleosynthesis. Whether this framework can accommodate the full range of observational constraints remains an open question.

In \cite{Shamir}, a revised version of the old "tired light" hypothesis \cite{Zwicky} is considered, in which redshift arises from photons losing energy due to the rotation of galaxies. Although the rotational effect might be real, the majority of the astrophysical community considers the tired light theory to be obsolete, having been ruled out by several independent lines of evidence \cite{TL1, TL2, TL3}. That said, the dismissal of the theory may itself be redshift-dependent, and we also note that the proposed connection between redshift and rotation in \cite{Shamir} appears rather superficial. Finally, we note that \cite{Twin} reports the observation of a spiral galaxy similar in structure to the Milky Way, already in existence just one billion years after the Big Bang. This discovery adds to the tension surrounding the age problem, as it had previously been assumed that the formation of such galaxies required several billion years of evolution.
 
\section{Conclusions}
The recent fits of cosmological parameters by the DESI collaboration \cite{DESI}, if confirmed, have the potential to shift our cosmological
model from $\Lambda$CDM to one that allows for a variable (scale factor-dependent) parameter $w(a)$ in the equation of state. This model would
most likely correspond to a quintessence model or $f(R)$
gravity. Given the significance of such a paradigm shift in cosmology,
we have conducted various  consistency checks. They include an
examination of the acceleration of the expansion, which has dominated
cosmology over the past few decades, with observational evidence
pointing to a positive acceleration and the theoretical efforts to
explain it. The DESI fit in the Quintessence parametrization shows a positive
acceleration, i.e. $\ddot{a} >0$,  roughly for $0<z<0.85$. But the
decelaration parameter at the present epoch does not agree with model
independent estimates. Nevertheless one could use the fact that in the
DESI-Quintessence model the acceleration lasts longer into the past as
a discriminating point while observing distant supernovae. The $\Lambda$CDM model has a positive
acceleration in the future ($a>1$) where the DESI-Quintessence model
shows here a negative value.

Another crucial cosmological parameter is the lifetime of the Universe.
Accepting the existence of old stars and galaxies, any viable model
should eventually be properly
constrained by these values. The DESI-Quintessence model satisfies this
requirement only for some values of the allowed parameter space. This
is true for the time dependent fit of the equation of state as well as
for constant one which would correspond to the $\Lambda$CDM model. The
problem of luminous objects at a high cosmological redshift reveals a
problem for both models which can be framed as a ``too early structure
formation'' after the Big Bang \cite{Steinhardt1, Laursen, Steinhardt3}. Since the parameter space of the DESI
fit allows also bigger lifetimes of the Universe we see this as an opportunity for the
time dependent DESI if, using an adequate time scale contraint from
structure formation, we allow only values leading to a bigger lifetime.

A remark is deemed necessary here to clarify the DESI-Quintessence terminology. As mentioned in Section~\ref{Sec2}, it is sufficient to consider the phenomenological equations (\ref{F1DE}) and (\ref{F2DE}), which might encompass a broader class of models-though they would likely be  indistinguishable from the Quintessence model, as the Friedmann equations are identical. Finally, our study is supplemented by numerical and approximate analytical solutions for the scale factor.

\end{document}